\documentclass[prd,aps,twocolumn,floatfix,nofootinbib]{revtex4}
\usepackage{amssymb,graphicx}
\usepackage{epsfig}
\usepackage[usenames]{color}

\newcommand{\beqn}{\begin{eqnarray}}
\newcommand{\eeqn}{\end{eqnarray}}
\newcommand{\beq}{\begin{equation}}
\newcommand{\eeq}{\end{equation}}

\begin{document}

\title{Eccentric black hole-neutron star mergers: effects of black hole spin and equation of state}
\author{William E.\ East${}^1$, Frans Pretorius${}^1$, and Branson C.\ Stephens${}^2$}
\affiliation{
${}^1$Department of Physics, Princeton University, Princeton, New Jersey 08544, USA. \\
${}^2$Center for Gravitation and Cosmology, University of Wisconsin-Milwaukee, Milwaukee, Wisconsin 53211, USA.
}

\begin{abstract}
There is a high level of interest in black hole-neutron star binaries, not only because 
their mergers may be detected by gravitational wave observatories in the coming years, 
but also because of the possibility that they could explain a class of short duration 
gamma-ray bursts. We study black hole-neutron star mergers that occur with high 
eccentricity as may arise from dynamical capture in dense stellar regions such as nuclear 
or globular clusters.  We perform general relativistic simulations of binaries with a 
range of impact parameters, three different initial black hole spins (zero, aligned
and antialigned with the orbital angular momentum), and neutron stars 
with three different equations of state. We find a rich diversity across these parameters in the 
resulting gravitational wave signals and matter dynamics, which
should also be reflected in the consequent electromagnetic emission.
Before tidal disruption, the gravitational wave emission
is significantly larger than perturbative predictions suggest for periapsis distances
close to effective innermost stable separations, exhibiting features reflecting
the zoom-whirl dynamics of the orbit there. Guided by the simulations, we develop
a simple model for the change in orbital parameters of the binary during close-encounters.
Depending upon the initial parameters
of the system, we find that mass transfer during nonmerging close-encounters can range
from essentially zero to a sizable fraction of the initial neutron star mass.
The same holds for the amount of material outside the black hole post-merger,
and in some cases roughly half of this material is estimated to be unbound.
We also see that nonmerging close-encounters generically excite large oscillations 
in the neutron star that are qualitatively consistent with $f$-modes.
\end{abstract}

\maketitle

\section{Introduction}
\label{intro}

It is important to understand mergers of black holes (BHs) and neutron stars (NSs)
not only because they are a chief source for
ground-based gravitational wave (GW) detectors (such as LIGO~\cite{LIGO})
but also because they may be accompanied by a diverse range of electromagnetic (EM)
and neutrino emission. One of the more interesting possibilities in this
regard is that BH-NS mergers may be progenitors of a fraction of observed short-hard 
gamma-ray bursts (sGRBs)~\cite{npp92,2005Natur.437..851G,2005ApJ...630L.165L}. 
Furthermore, existing and planned wide-field
survey telescopes such as PTF~\cite{2009PASP..121.1334R}, Pan-STARRS~\cite{2004SPIE.5489...11K}, 
and LSST~\cite{2009arXiv0912.0201L}
are beginning to observe and classify fainter EM transient events,
some of which are expected to be produced in BH-NS mergers from a variety of mechanisms
(see \cite{Metzger:2011bv} for a recent, detailed exploration of the possibilities).
Coincident observation
of these EM events with a gravitational wave signal from a binary coalescence 
would provide a wealth of additional information about the system beyond
any individual observation, even, for example, providing an independent way to measure
cosmological parameters~\cite{dalal}.
However this depends on having a good understanding of
how to map the particular observations to the underlying astrophysical
processes governing the merger.

The great diversity in sGRBs hints at the possibility that there will be a
corresponding diversity in the associated GW and EM 
signals, and motivates the exploration of all viable channels
for binary compact (BCO) mergers.
BCOs may form through the evolution
of primordial binaries or through dynamical capture in dense stellar
systems, such as nuclear or globular clusters.
BCOs formed through the latter channel, which could merge with non-negligible eccentricity,
are the focus of this work, which is a follow-up of a first study of
dynamical capture BH-NS evolution using full general relativistic 
hydrodynamics~\cite{ebhns_letter}. 
Before delving into the details, we briefly summarize the motivation for studying this class of BCO; 
see~\cite{ebhns_letter} and a related study on eccentric NS-NS mergers~\cite{2011arXiv1109.5128G}
for more discussion.  In contrast, most studies of BH-NS mergers to date have focused on the 
quasicircular inspiral case; see~\cite{lrr-2011-6} for a review of these efforts.

A chance close-encounter of a BH and NS in a stellar cluster could result in a bound
system if the energy loss due to GW emission is sufficiently large. (Tidal interaction with
the NS is also a source of energy loss~\cite{PressTeuk}, though due to the relative scalings with distance, 
to leading-order the total cross section for {\em capture} is dominated by the GW emission.) 
Because of gravitational focusing, a sizable fraction of binding encounters will result in highly eccentric binaries that merge
within a few orbits of the initial encounter.
For Newtonian hyperbolic orbits of systems with total mass $M$ and relative velocity $w$ at infinity, the pericenter 
separation $r_p$ is related to the impact parameter at infinity $b$ by $r_p = b^2w^2/2M + O(w^4)$
(unless otherwise stated we employ geometric units with $G=c=1$ throughout).  Thus,
since the capture rate is linearly proportional to the cross sectional area, it is linearly proportional to $r_p$.
For the 4:1 BH to NS mass ratio systems studied here (with $w=1000$~km/s), estimates of the energy
lost to GW based on the work of~\cite{pm63} shows binding encounters occur at $r_p \lesssim 40$ M (and
scales with mass ratio as $\mu^{4/7}$).
The results presented here show that cases with initial $r_p \lesssim 7M$ (for the canonical nonspinning BH), or
close to $20\%$ of binding encounters (of a 4:1 system with $w=1000$~km/s), result in ``direct collisions,'' {\em i.e.} 
merging on the first close-encounter.

At first glance it may seem that such encounters would be too rare to be of any relevance
as possible GW or EM transient sources; however, recent studies have suggested otherwise.
In~\cite{oleary} estimates of BH-BH encounters in galactic nuclear clusters
suggest Advanced LIGO rates of O($1-10^3$) per year, while 
in~\cite{lee2010} (following earlier work in~\cite{grindlay2006,2008MNRAS.388L...6S,2009A&A...498..329G})
it was argued that direct capture NS-NS mergers in globular clusters could
account for a sizable fraction of observed sGRBs. In this latter work,
a procedure is also provided for scaling the NS-NS results to BH-NS systems,
which yields BH-NS dynamical capture event rates of O($10-10^2$) yr$^{-1}$ Gpc$^{-3}$ 
(using an estimate of the relative fraction of BH to NSs $f_{\rm BH}/f_{\rm NS}\approx 0.28$
as suggested in~\cite{lee2010}).
In comparison, population synthesis models~\citep{belczynski} find primordial
BH-NS merger rates from
$\sim0.1~$yr~${}^{-1}$Gpc${}^{-3}$ (pessimistic) to
$\sim120~$yr~${}^{-1}$Gpc${}^{-3}$ (optimistic).
To estimate Advanced LIGO detection rates would require full templates for
these events, which we relegate to a future study; however, in~\cite{Kocsis:2011jy}
the signal-to-noise ratio was computed using post-Newtonian-based
models of the early stages of the mergers, which showed that a subset (depending
upon the component masses) could be observed out to 200-300 Mpc for an average orientation, even excluding 
the final stages of the merger in the templates. This suggests detection rates of $\sim 0.3 - 10$/yr.
For comparison to primordial BH/NS binaries, \cite{2010CQGra..27q3001A}
quotes an Advanced LIGO optimal detection distance of $927$ Mpc for a
10 M$_{\odot}$ BH to 1.4 M$_{\odot}$ NS merger, which scales to $\sim 400$Mpc
when averaging over orientation and sky location; the corresponding detection
rates range from $0.2$/yr to $300$/yr for pessimistic to optimistic source population estimates.

Though far from conclusive, there is also observational evidence for multiple
sGRB progenitors, which could, in part, be due to dynamical capture vs
primordial mergers. Of sGRBs with identified host galaxies, $\sim 25$\% 
have offsets of $\gtrsim 15$~kpc from their
hosts~\cite{berger2010}.  This subset of sGRBs with large offsets would be
consistent with kicked, primordially formed BCOs or with dynamically formed
binaries in globular clusters.  The latter may be preferred for the largest
offsets~\cite{church}, especially if primordial BCOs experience weak
kicks~\cite{dewi}.
Analysis of x-ray afterglows observed by Swift/X-Ray Telescope
suggests that different progenitors may be responsible for sGRBs with and without extended 
emission~\cite{Norris:2011tt}; again, one possible
explanation is dynamical capture (with extended tidal tails leading to 
long-term emission) vs primordial.
There has also been
a claim that a very high-energy gamma-ray
source observed in Terzan 5 may, in fact, be the remnant of
a BCO merger-powered sGRB~\cite{Domainko:2011gv}; if true, this would support
the claim that dense cluster environments can be significant sources
of BCO mergers.

It has also been suggested that some merging NS-NS or BH-NS binaries
are actually part of a coeval or dynamically formed triple~\cite{Thompson}.  The tertiary (for example
a white dwarf) drives the binary to high eccentricities through a Kozai resonance, resulting
in a faster merger time.
Though dynamically different from direct capture two-body systems, in Kozai-accelerated evolution the
merger itself could take place with significant eccentricity~\cite{Wen:2002km} and have comparable 
behavior at late times to the systems studied here. However, we are unaware of 
any systematic studies of expected populations and corresponding LIGO-event rates
for such Kozai triple systems.

Merging with moderate to high eccentricity could have significant effects
on all the GW and EM observables from the event. In contrast to a low-eccentricity
inspiral, the GWs are emitted primarily around periapsis, resulting
in waveforms that resemble a sequence of bursts more than a continuous 
signal. Consequently, for mass ratios relevant to stellar mass BH-NS mergers, the evolution of 
effective orbital parameters describing the binary is not adiabatic, {\em i.e.}
it does not occur quasistatically as in the early inspiral of low-eccentricity binaries.
Regarding effects associated with tidal disruption of the NS,
an interesting coincidence for this class of BH-NS systems is that 
the typical radii inside of which tidal stripping begins, depending upon the NS compactness 
and, hence, equation of state (EOS), 
roughly coincides with the
range of pericenter separations where the orbit becomes unstable 
due to general relativistic effects ($r_p\lesssim 10M$ depending upon
on the spin of the black hole where $M$ is the total mass of the system).
The resultant dynamics could thus be very different from Newtonian expectations.
Of course, these two zones also roughly correspond for quasicircular inspiral.  
However, highly eccentric binaries have significantly more
angular momentum (at a given orbital separation) than quasicircular binaries.
This additional angular momentum could strongly affect the matter dynamics
relative to the quasicircular case, for example, resulting in comparatively massive
disks and/or in multiple and prolonged episodes of mass transfer and ejection. 
Ejected mass will decompress and form heavy elements through
the r-process~\cite{1974ApJ...192L.145L,Rosswog:1998gc,Li:1998bw}, and thus these systems could 
account for a significant fraction of such elements in the Universe. Furthermore,
subsequent decay of the more radioactive isotopes could lead
to observable EM counterparts~\cite{2010MNRAS.406.2650M,Metzger:2011bv}.

In \cite{ebhns_letter} we presented the first simulations of BH-NS hyperbolic 
capture and merger for a nonspinning BH and a single equation of state,
modeled using general relativistic hydrodynamics (GRHD) (see~\cite{lee2010}
for an earlier related study using a Newtonian-based hydrodynamics code).
We found a striking dependence of the outcome---disk mass,
unbound material, GW signal---on the impact parameter. 
However, astrophysical BHs are expected to form with a range of spins,
(see e.g.~\cite{Gammie:2003qi,Miller:2011jq})
and in the quasicircular case spin was found to be crucial in obtaining 
significant accretion 
disks~\cite{illinoisBHNSspin, Foucart:2010eq,Kyutoku:2011vz,Chawla:2010sw}.
Additionally, there is significant uncertainty about the NS equation of state.  Since the EOS
determines the NS compaction and, hence, the point at which the NS will become 
tidally disrupted, it is also an important determinant of the merger outcome.
Here, we expand upon the previous analysis and consider initial BH-spins aligned
and antialigned with the orbital angular momentum (with dimensionless spin parameter
$a=0.5$ and $a=-0.5$, respectively), as well as different EOSs (the ``2H'', ``HB'' and ``B''
models from~\cite{jocelyn}). Note that there is no {\em a priori} reason to expect
alignment of the BH spin with orbital angular momentum in dynamical-capture binaries;
this particular choice of spin direction and magnitude was purely motivated 
as a first, simple exploration of the effects of spin on the merger. Certainly
in future studies a broader expanse of parameters will need to be considered. 
There is also much room for improvement with the matter description, including
more realistic EOSs and additional physics beyond GRHD.
For quasicircular BH-NS inspiral~\cite{Shibata:2006ks}, EOS effects were studied
in~\cite{Duez:2009yy,Pannarale:2011pk,shibataBHNS3,Lackey:2011vz}, the effects of
magnetic fields in~\cite{chawla, 2011arXiv1112.0568E}, and higher mass ratio systems (up to 7:1) in~\cite{Foucart:2011mz}.

In the remainder of the paper we begin with a brief review of our numerical methods (Sec.\ref{methods});
more details are given in a companion paper~\cite{East:2011aa}.
In Sec.~\ref{cases}, we describe the particular cases we study. We focus
on systems with small initial periapsis $r_p$, in part for the practical reason
that these binaries merge quickly and are thus computationally tractable,
this is the regime where full general-relativistic (GR) effects will be most
strongly manifested, and because maximum complementary information to post-Newtonian studies 
(e.g.~\cite{Kocsis:2011jy} or ~\cite{lee2010}) can be obtained.
We present the results of the simulations in Sec.~\ref{results}.
We find that the rich variability in the dynamics and merger outcome
as a function of impact parameter is compounded by considering 
different EOSs and values of BH spin.  For example, we find that with 
prograde spin or a stiffer EOS significant episodes of mass transfer may
occur during nonmerger close-encounters.  Systems where the BH has retrograde spin or
that are somewhat less eccentric (than parabolic) can undergo
sustained whirling phases before merger that are evident in the GW signal.
As in the eccentric binary NS mergers studied in~\cite{2011arXiv1109.5128G},
we also find that strong $f$-mode oscillations in the NS can be excited in 
close, nonmerger encounters (see also an earlier study of a head-on BH-NS
collision~\cite{Loffler:2006wa}, though here the presence of the
f-mode is largely a consequence of the initial data).
In Sec.~\ref{sec_evo_orb} we use the results of the simulations to calibrate
a simple model for the nonadiabatic evolution of effective orbital parameters
during each close-encounter. This will be used in a future study to
build model GW templates and explore the detectability of these systems
with gravitational wave observations. We conclude in Sec.~\ref{conclusions}.

\section{Numerical methods}
\label{methods}

In this section, we briefly outline our numerical methods for solving the GRHD
equations. More details are presented in \cite{East:2011aa}, including 
results from an extensive range of tests.

\subsection{Evolution}
We model the evolution of BH-NS binaries by solving the Einstein field equations coupled to 
a perfect fluid.  We solve the field equations in the generalized harmonic 
form~\cite{1985CMaPh.100..525F,garfinkle,gh3d} with constraint damping~\cite{Gundlach:2005eh,paper2},
where the coordinate 
degrees of freedom are specified through source functions $H^a$.
We employ fourth order
accurate finite difference (FD) techniques and Runge-Kutta time-stepping.
In order to avoid singularities we excise within any apparent horizons.
In evolving eccentric binaries,
we find that a damped harmonic gauge similar to the one described 
in~\cite{ultra_rel} (see also~\cite{damped_harmonic}) is 
beneficial for achieving stable evolutions through merger.
Specifically a damped harmonic gauge takes the form $H^a=\xi (n^a-\bar{n}^a)$, where $n^a$ is the 
four-vector normal to the constant coordinate time slices, $\bar{n}^a$ is another timelike unit vector, 
and $\xi$ is a constant controlling the magnitude of the damping.   
The particular form for $\bar{n}^a$ that we found to work well is from ~\cite{damped_harmonic},
\begin{equation}
\bar{n}^a=\frac{1}{\alpha}\left(\frac{\partial}{\partial t}\right)^a+\log\left(\frac{\alpha}{h^{1/2}}\right)n^a .
\end{equation}
We use a value of $\xi \approx 0.2/M$, though include a spatial dependency so that $\xi$ goes to zero at 
spatial infinity.  In addition, we begin with initial data in the harmonic gauge and transition to this
damped harmonic gauge before the two objects begin to strongly interact.  The use of this gauge seems
to smooth out sharp features in the lapse that develop near merger when harmonic gauge is used,
however, we have not studied the pure harmonic case in sufficient detail to conclude whether
or not harmonic time slicing is developing a coordinate singularity.

We describe the neutron star material as a perfect fluid.
The fluid equations are written in conservation-law form and solved using high-resolution shock-capturing
schemes.  Though we have implemented several methods for calculating intercell fluxes
and for reconstructing fluid primitive variables at cell faces,
we used HLL~\cite{hll} combined with WENO-5~\cite{weno5} for the results presented here.
The fluid is evolved with outflow boundary conditions using second-order Runge-Kutta time-stepping. 
To resolve the various length scales, we use Berger- and Oliger- style adaptive mesh refinement (AMR)~\cite{bo84}.
AMR boundaries require special treatment in conservative hydrodynamics codes, since 
the fluxes emerging from a fine-grid region, for example, do not exactly match the 
incoming flux calculated on the coarse grid due to their differing truncation errors. 
To enforce conservation, we correct the adjacent coarse-grid cells using the fine-grid 
fluxes, according to the method of Berger and Colella~\cite{bc89};
more details on our particular implementation of this method is given in~\cite{East:2011aa}.

Truncation error estimates are used to generate the AMR-level structure. All initial
data were evolved with a fiducial ``medium'' resolution run, where the coarsest-level 
has $128^3$ cells and covers the entire domain (we use a compactified coordinate system, so this
includes spatial infinity). We also chose a maximum truncation error so that 
initially six additional levels of refinement (seven total) are generated to resolve the BH and NS. Given the
computational expense of these simulations, we limited the total number of levels to
seven during evolution, and did not allow regridding on the two coarsest-levels to prevent
the algorithm from tracking the outgoing gravitational waves beyond the 
largest extraction sphere of $100M$.
For several representative cases we also ran a ``low'' and ``high'' resolution
simulation for convergence testing, where on each level the low (high) resolution
run had a mesh spacing of $64/50$ ($64/96$) of the medium resolution run,
and we scaled the corresponding maximum local-truncation error threshold-parameter 
used by the AMR algorithm assuming second-order convergence. Note that this
procedure will not generate identical hierarchies between different resolution
runs (except for the two coarsest levels in the wave zone, which we keep fixed),
but on average the highest-resolution grid covering a given coordinate cell will have
the above refinement ratios between the different runs.
To give some sense of the smallest scales resolved by the hierarchy, before tidal disruption, 
the low (medium, high) resolution
run has two finest-level meshes centered around the BH and NS of roughly $80^3$ ($100^3$, $150^3$)
cells each, resolving the diameter of the NS with approximately $40$ ($50$, $75$) cells
and the BH horizon diameter with roughly $70$ ($85$, $130$) cells.
Unless otherwise noted, results will be reported for medium resolution with
error bars (where appropriate) computed from convergence
calculations.

\subsection{Initial data}
\begin{figure}
\begin{center}
\includegraphics[width=3.2in,clip=true,draft=false]{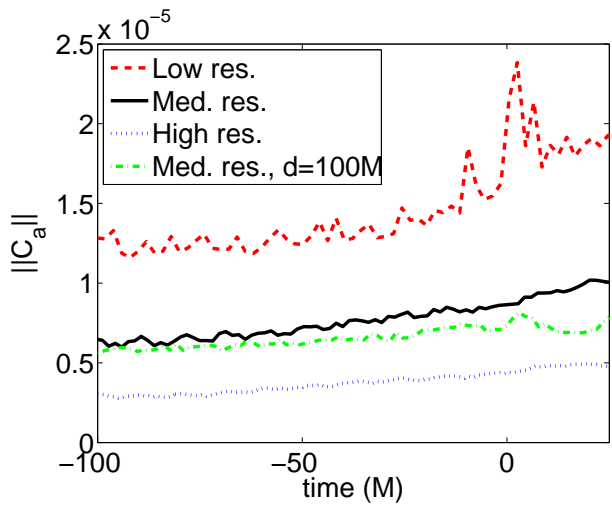}
\end{center}
\caption{
$L^2$-norm of the constraint violation, $C_a \equiv H_a-\Box x_a$, in units of $1/M$ from the $r_p=10M$ case in the 
$100M \times 100M$-region around the center of mass in the equatorial plane ({\em i.e.} $\sqrt{\int\|C_a\|^2 d^2x / \int d^2x}$ ). 
This is shown for low, medium, and high resolutions 
for the standard initial separation between the BH and NS of $d=50M$;
the relative magnitudes at a given time are consistent to good approximation
with second-order convergence.  
Also shown is a medium resolution run with an initial separation of $d=100M$.  The time is shifted so that the 
point of closest approach occurs at approximately $t=0$ for all cases.
}
\label{norm_cnst}
\end{figure}

We construct initial data by superimposing a boosted BH with a boosted nonspinning Tolman-Oppenheimer-Volkoff (TOV)
star solution separated by $d=50 M$.  Though this superposition does not strictly satisfy the constraint equations
except in the limit of infinite separation, we have performed tests at various separations in order to verify that 
the superposition-induced constraint violation is comparable to truncation error at our resolution, in particular
following an initially slightly larger transient that propagates away (and is partly damped due to the use of constraint damping)
on roughly the light-crossing time $d$ of the binary. 
In Fig.~\ref{norm_cnst},
we show the level of constraint violation ($C_a \equiv H_a-\Box x_a$) following the transient 
for various separations and resolutions for the $r_p=10M$ case.
At these resolutions, we can still achieve convergence of the constraints with this superposed data, and 
increasing the initial separation to $d=100 M$ does not significantly affect the level of constraint violation at the 
same resolution. This implies that the error introduced by the superposition is on the order of or smaller
than the numerical truncation error at this resolution. 

The fact that we are beginning the binary at finite 
separation as a simple superposition of boosted single-compact object solutions
and that there is a transient early time constraint-violation, both effectively introduce systematic errors in the parameters of the binary.
To give some idea of the possible magnitude of this error, the effect on the apparent horizon-mass of the BH at early times and 
the amplitude of the density oscillation induced in the NS are both $\lesssim 2\%$.
Though it is difficult to exactly quantify how this will translate to modified-binary parameters, 
we expect it to be comparable to or smaller than the error introduced by
setting initial orbital parameters at a finite distance based on a Newtonian
approximation (as described in the next section).
As we report later, the truncation error in quantities of interest that we
do measure such as energy emitted in GWs is percentwise larger
than this, implying that solving the constraints or attempting more
accurate initial representations for the metric and NS fluid distribution may only offer marginal improvement 
in the overall accuracy of the results at these resolutions. However, for future higher-resolution studies it 
would be important to solve the constraints and improve the model of the physical initial conditions for the system.

Finally, we briefly comment that since our simulations employ compactified coordinates 
such that the outer boundaries extend to spatial infinity, the global [Arnowitt-Deser-Misner (ADM)]
$M$ and $J$ should be conserved.  In practice, however, we must
evaluate these quantities at a finite distance, making them
subject to gauge artifacts, some propagating outward from the central BH-NS region from $t=0$.
For $t<200M$, an extraction sphere of $300M$ is free of propagating artifacts, hence
$M$ ($J$) is conserved to better than $0.3$ (2.0)\% for all cases at medium resolution.

\section{Cases}
\label{cases}
Motivated by possible BH-NS interactions in cores of galactic nuclei and
globular clusters, we create initial data for hyperbolic encounters with
varying impact parameter.  These encounters are hyperbolic
since the two bodies have nonzero kinetic energy at (effectively)
infinite separation.  
We will take their relative velocity at infinity
to be $w=1000$~km/s since this is the expected magnitude of the virial
velocity in the core of a nuclear cluster~\cite{oleary,qs87}.

In practice, this 
initially positive total energy is small compared to the kinetic energy
of the encounter itself, so that the orbits are nearly parabolic and 
have eccentricities $1+O(10^{-5})$.  In this study we  also restrict our attention to BH-NS systems with a 
4:1-mass ratio (referring to the isolated ADM masses of the BH and NS). 
The reasons for choosing this mass ratio are in part because it is within the
range of astrophysically plausible values given current observations of NS and candidate 
BH masses (see, for example,~\cite{Lattimer:2010uk,Ozel:2010su}), and in part because within this range it 
is also a value where we expect to see strong tidal-disruption effects. Certainly it would be of interest
to explore a broader range of mass ratios, however, due to limited computational
resources we leave that to a future study.

We choose the initial positions and velocities for the BH and the NS 
according to the Newtonian equations for a hyperbolic orbit (see, e.g., 
\cite{turner77}) with $w=1000$~km/s and an initial separation of $d=50M$.  
We vary the initial spin of the BH, considering the values $a=-0.5,0$, and $0.5$ where negative (positive) 
values indicate retrograde (prograde) spin in relation to the orbital angular momentum.  
We also consider three different broken $\Gamma$-law model EOSs labelled 2H, HB, and  B in \cite{jocelyn}.
For the prototypical 1.35 M$_{\odot}$ NS that we use, these EOSs give compactions $M_{NS}/R_{NS}=0.13,0.17$, and $0.18$, respectively.   
Note that while the compaction of the B EOS NS is only slightly smaller than that of the HB, it has a maximum mass of 
2.0 M$_{\odot}$ (the 2H and HB have maximum masses of 2.83 and 2.12 M$_{\odot}$, respectively).  
Given recent observations \cite{Demorest}, B is therefore on the soft end of the allowed range for the EOS family considered here. 
We include a thermal component in the EOS, a $\Gamma$-law with $\Gamma_{\rm th}=1.5$ to allow for shock heating.

We do not consider large impact parameters corresponding to initial
$r_p>15M$. Even so, for impact parameters at the upper end of the range
we do evolve the eccentricity is sufficiently large (though $<1$) after 
the first close-encounter that it would be very expensive to evolve
to the second encounter. To help calibrate our model for orbital parameter
evolution that will be introduced in Sec.~\ref{sec_evo_orb}, we
also consider a set of runs with initial orbital parameters for a 
bound orbit with eccentricity $e=0.75$ instead of the hyperbolic orbit with $e\approx1$.  
These simulations can be seen as corresponding to systems that have already undergone one or 
more close-encounters and evolved to these orbital parameters.

\begin{figure*}
\includegraphics[width=6.2in,clip=true]{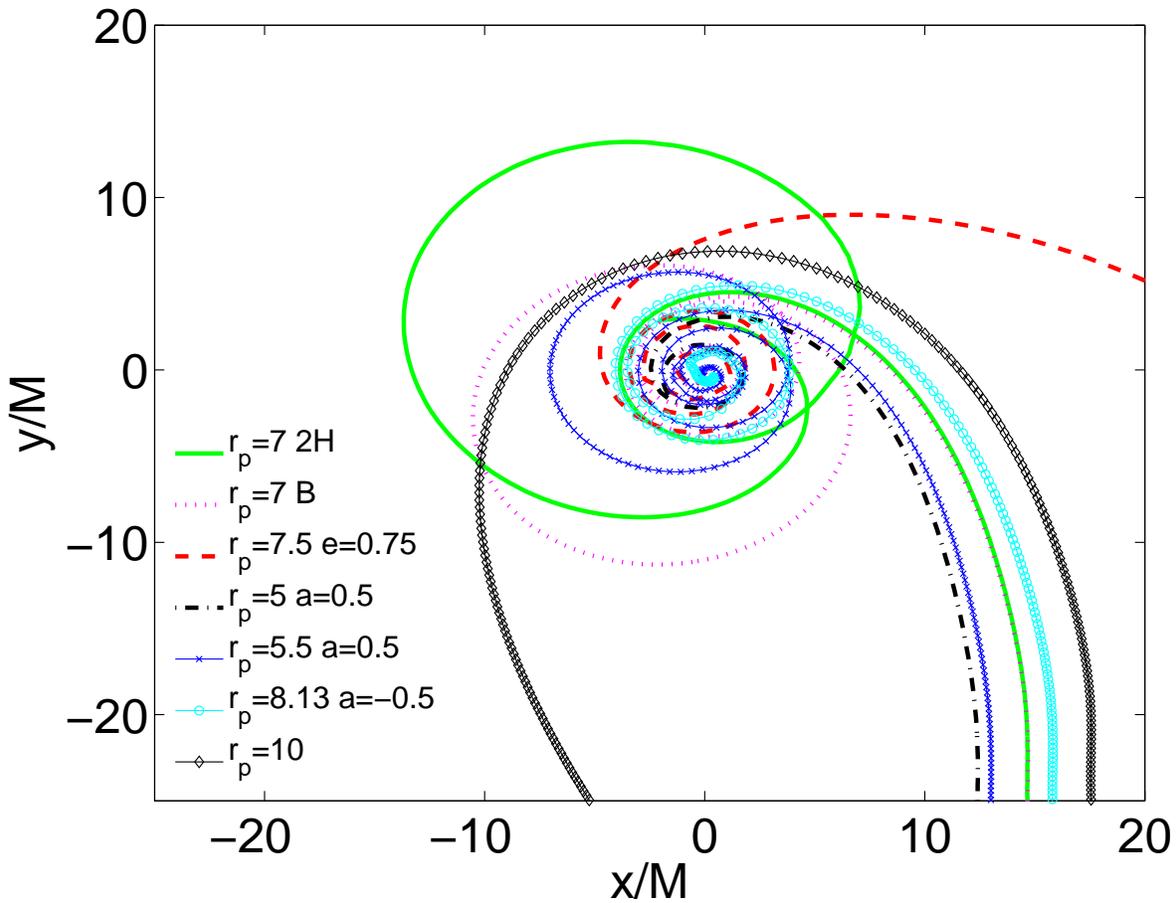}
\caption{
Trajectories of the NS center-of-mass for various simulations.
The $r_p=7.0$ with 2H EOS, $r_p=7.0$ with B EOS, and $r_p=5.5$ $a=0.5$ (HB EOS) both undergo
a close-encounter followed by a short elliptic orbit before merging.  Note that while in both the
$r_p=7.0$-cases the NS approaches the BH on essentially the same orbit, the dynamics around the close-encounter and subsequent orbits are very different due to EOS effects (see Sec.~\ref{sec_eos}).
The $r_p=10$ (HB EOS) undergoes a long-period elliptical orbit following the initial periapsis passage. 
The remainder of the cases shown merge on the first encounter while displaying various degrees of whirling
behavior.
}
\label{trajplot}
\end{figure*}

\begin{figure*}
\includegraphics[clip=true, draft=false, viewport=0 0 700 500, width=2.16in]{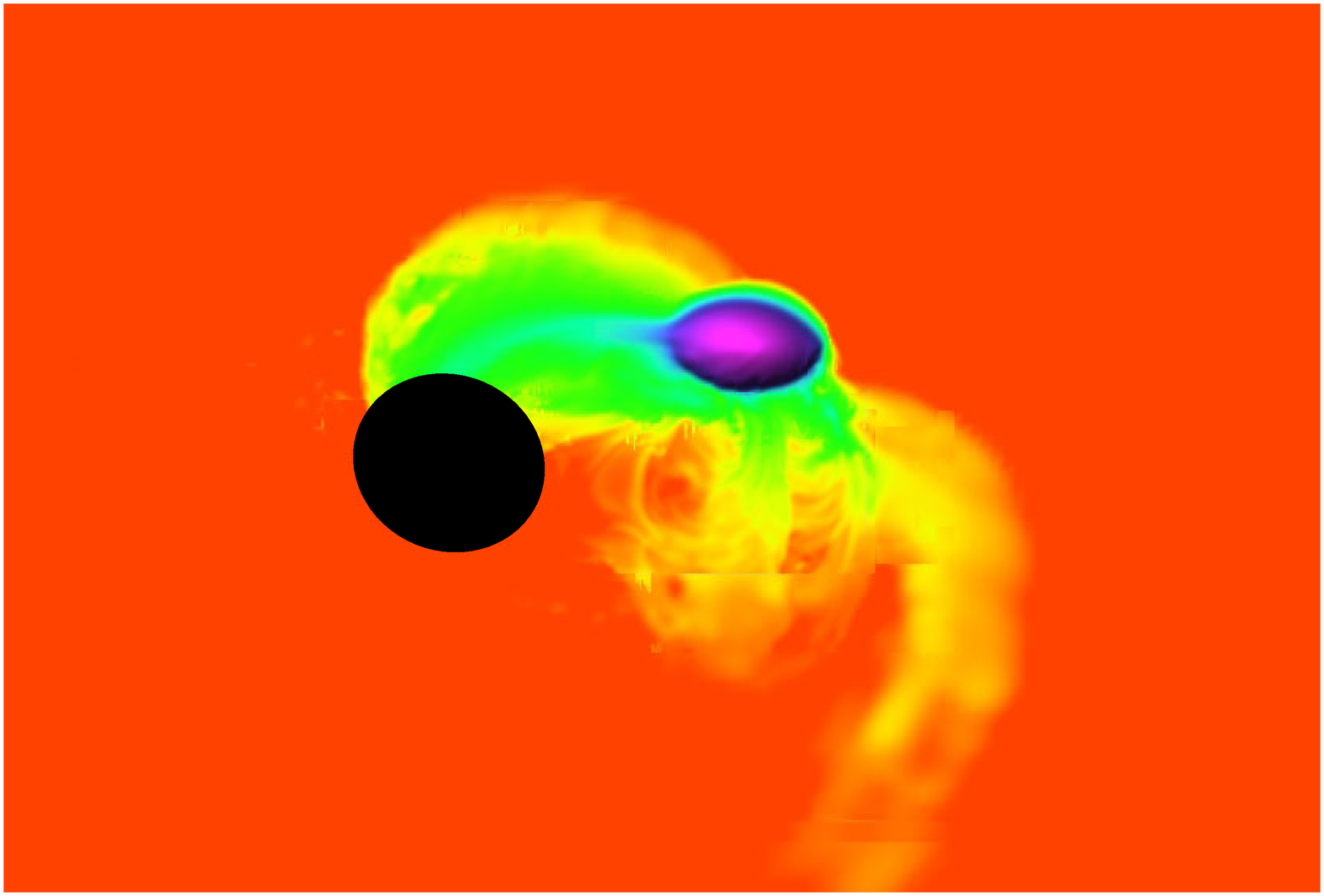} \hspace{0.01 in}
\includegraphics[clip=true, draft=false, viewport=50 20 750 520, width=2.16in]{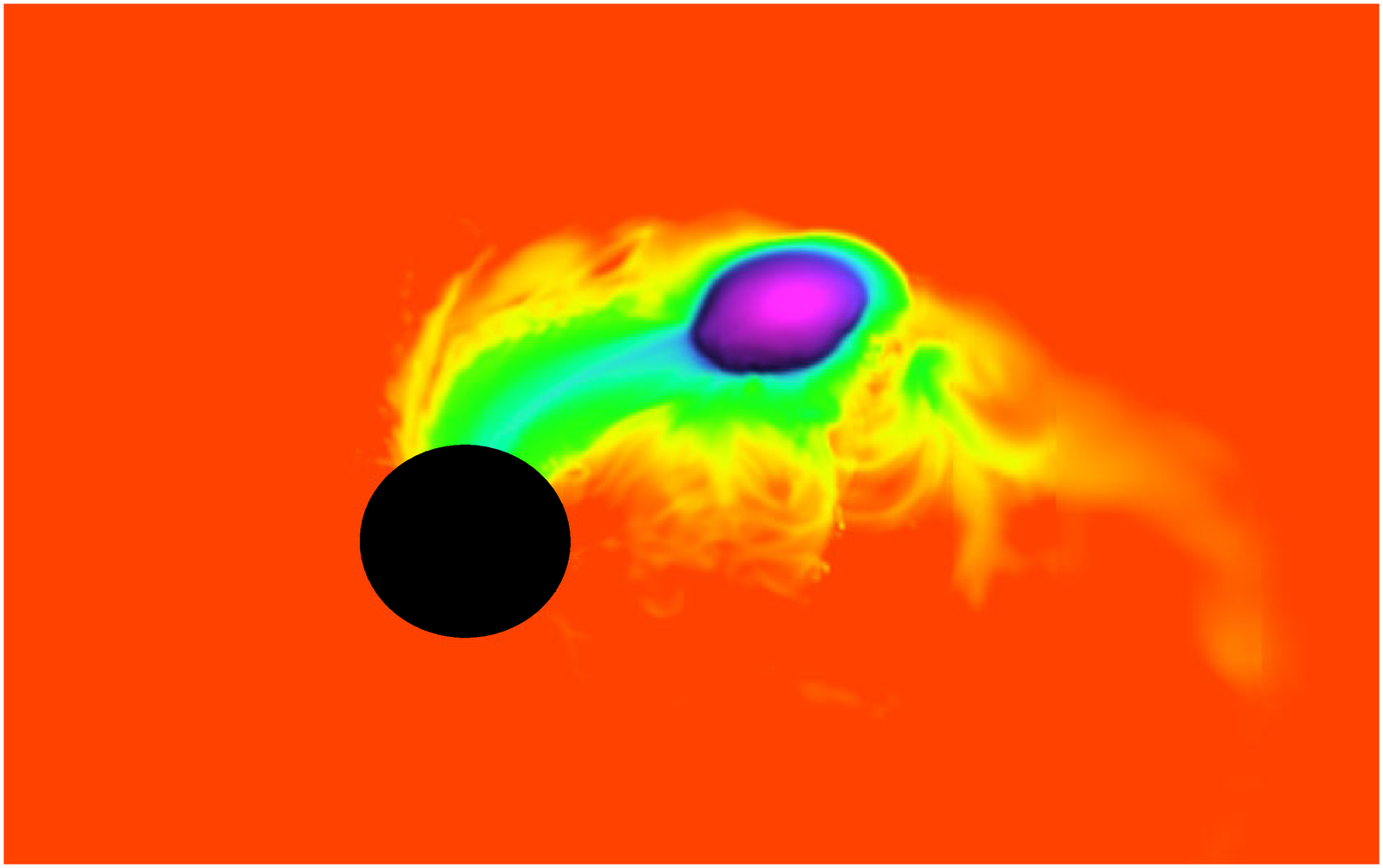} \hspace{0.01 in}
\includegraphics[clip=true, draft=false, viewport=50 0 750 500, width=2.16in]{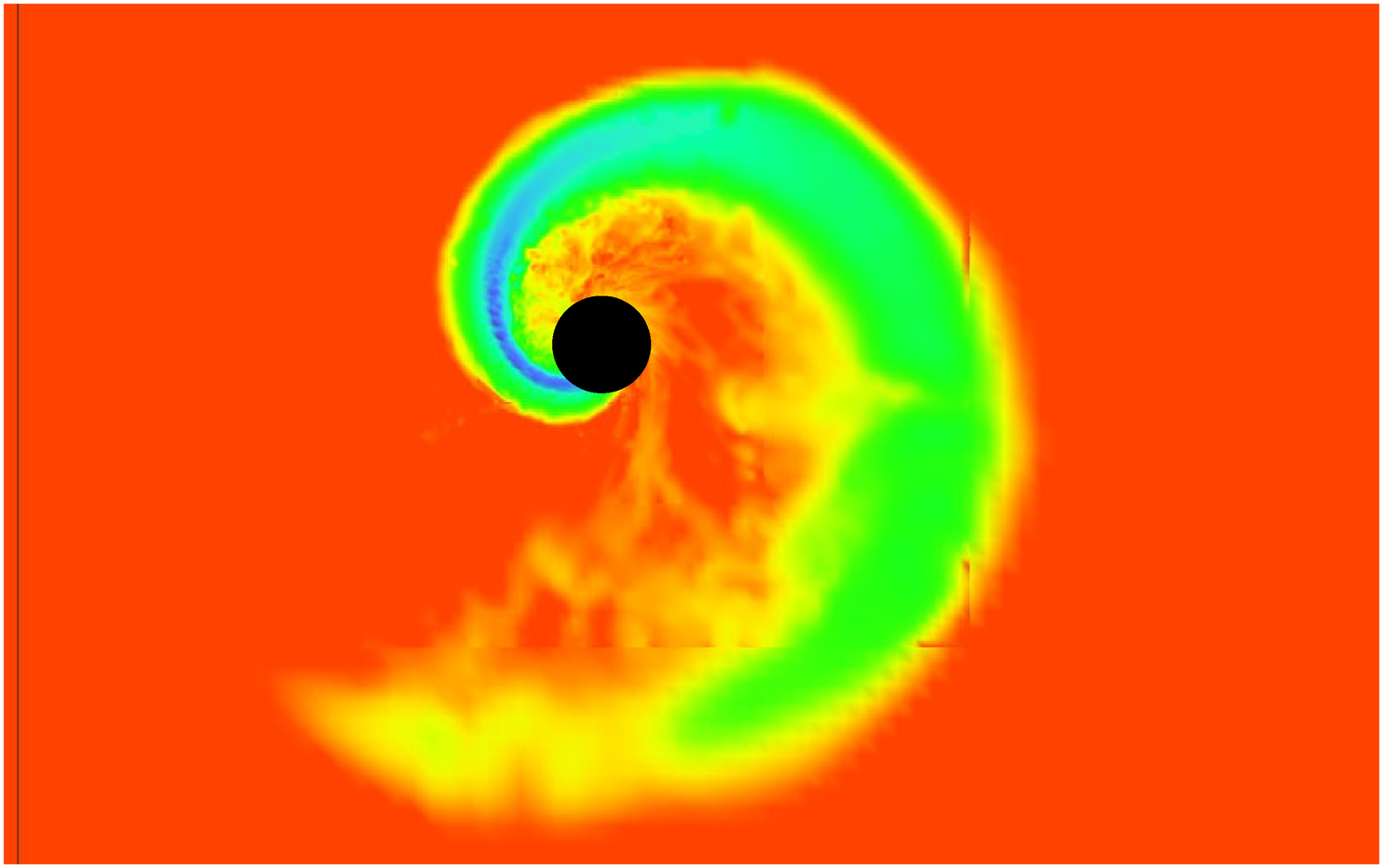} \\ \vspace{0.075 in}
\includegraphics[clip=true, draft=false, viewport=50 50 750 550, width=2.16in]{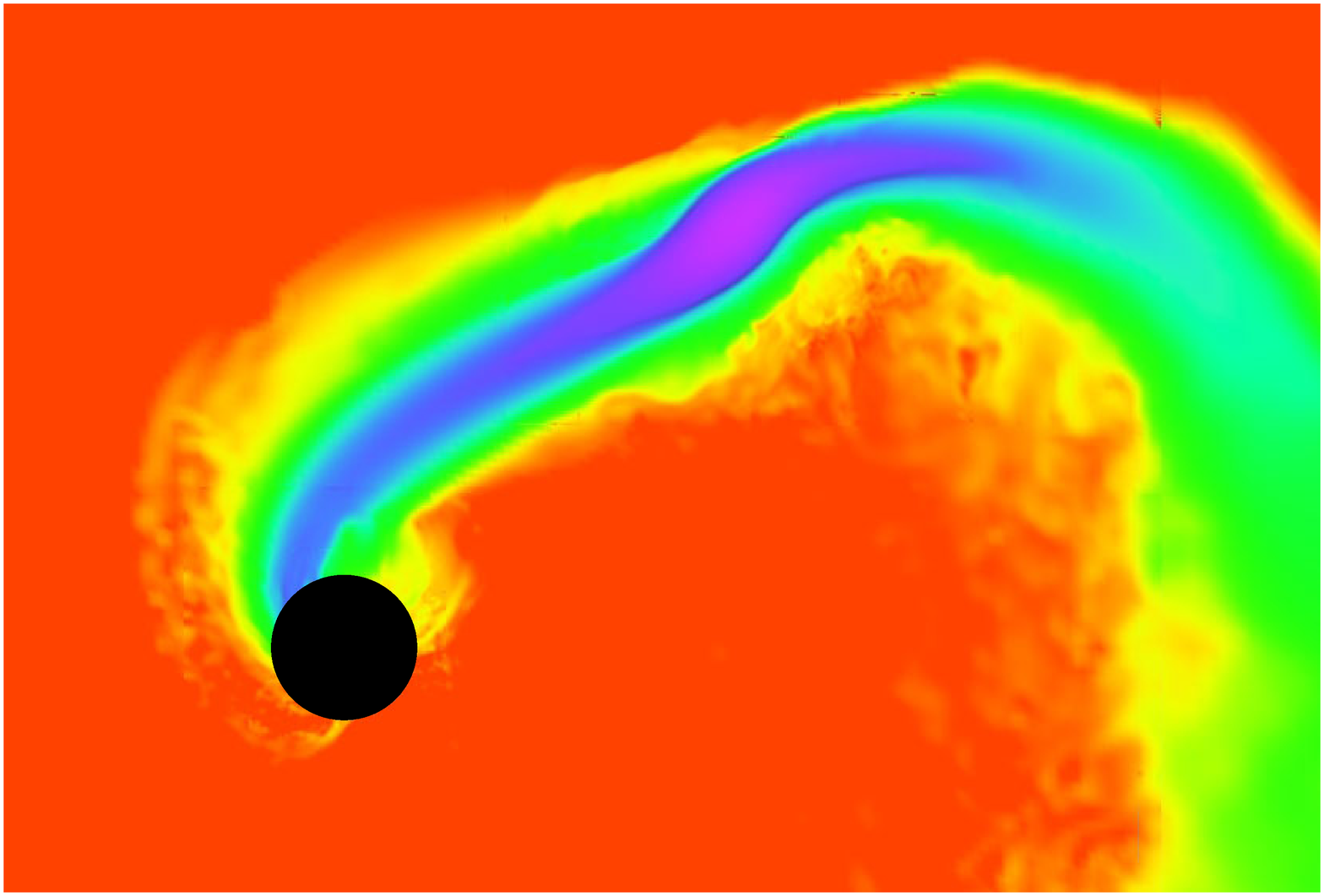} \hspace{0.01 in}
\includegraphics[clip=true, draft=false, viewport=90 0 790 500, width=2.16in]{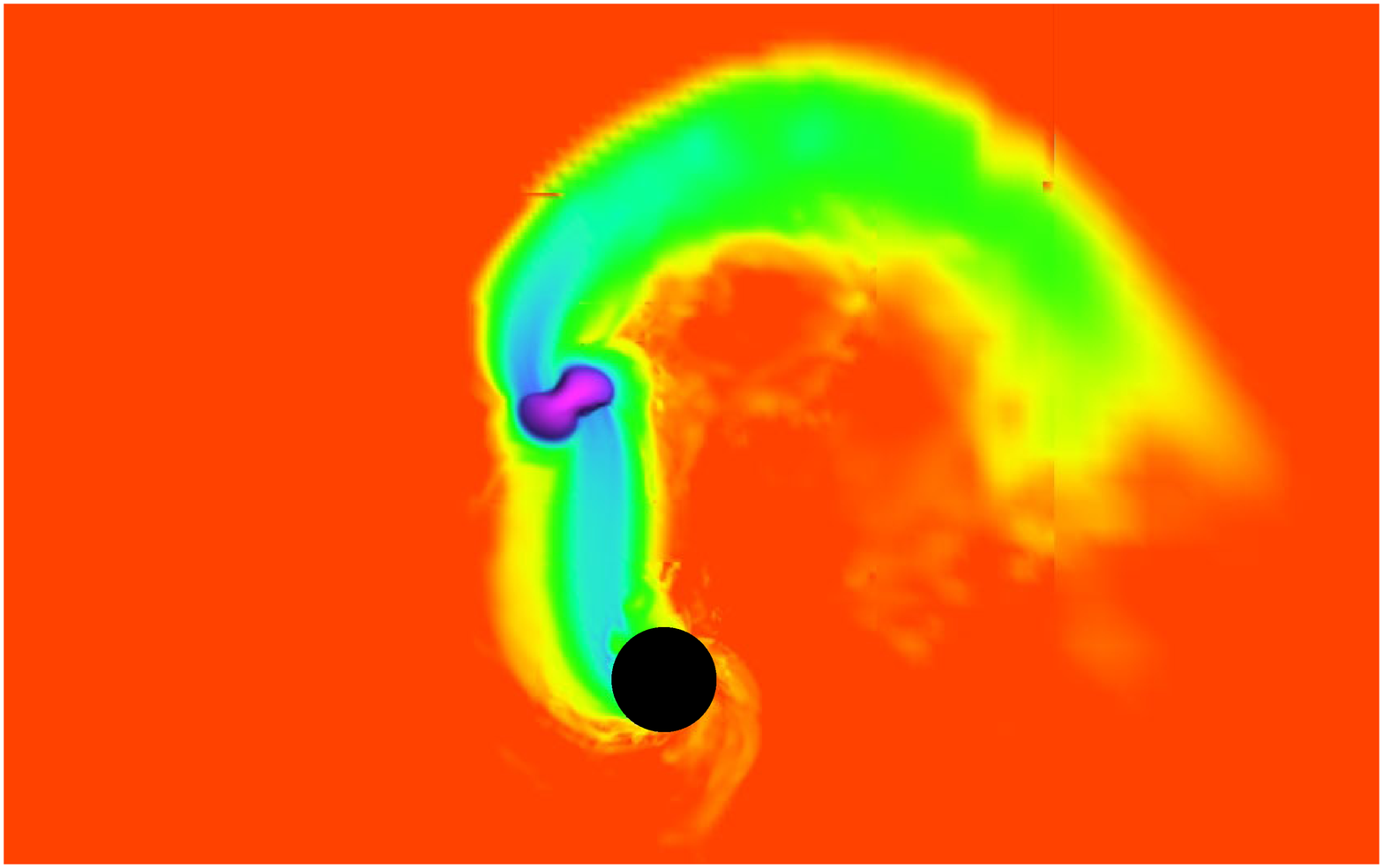} \hspace{0.01 in}
\includegraphics[clip=true, draft=false, viewport=0 30 700 530, width=2.16in]{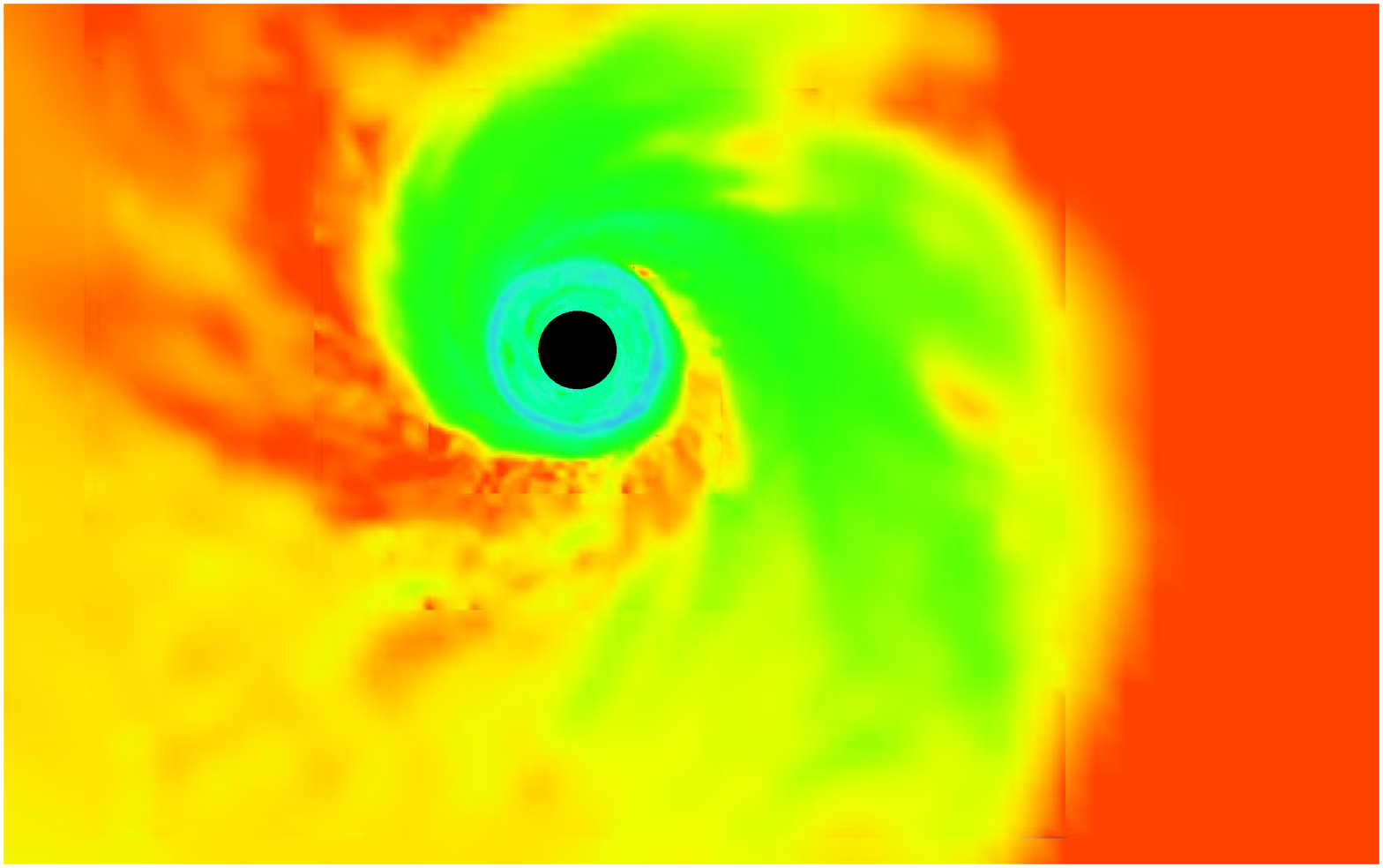} 
\caption{
Rest-mass density in the equatorial plane from various BH-NS simulations, left to right, top to bottom:   
(1) the BH and NS undergoing a close-encounter ($t=242$ M, $r_p=7.0$, B EOS), 
(2) the NS undergoing a whirling-phase before merging ($t=269$ M, $r_p=8.13$, $a=-0.5$), 
(3) the NS stretched into a long tidal stream during merger ($t=506$ M, $r_p=7.5$, $e=0.75$), 
(4) a mass transfer episode ($t=292$ M, $r_p=7.0$, 2H EOS),
(5) towards the end of a mass-transfer episode during the NS's first periapsis passage ($t=272$ M, $r_p=5.5$, $a=0.5$), and
(6) a nascent accretion disk ($t=388$ M, $r_p=5$, $a=0.5$).
Recall that $r_p$ is reported in units of 
total mass $M$.
The color scale is logarithmic from $10^{-6}$ to $1$ times the initial maximum density ($\rho_{\rm max}=$ 8.3 (3.7,9.8) $\times 10^{14}$ g cm$^{-3}$ for the HB (2H, B) EOS).
The BH is roughly the same coordinate size (with diameter $\approx3$ M) in all 
panels, which can be used
to infer the relative scale of each snapshot. 
}
\label{density}
\end{figure*}

To keep the parameter space at a manageable size, we vary only one of the three parameters of 
BH spin, NS-EOS, and initial eccentricity at a time from our base case --- an 
initially nonspinning BH ($a=0$), a NS with the HB EOS, and the two objects initially with Newtonian orbital parameters 
corresponding to a marginally unbound orbit ($e\approx1$) --- and then consider a range of impact parameters.  
See Fig.~\ref{trajplot} for plots of the NS trajectory for several cases, and Fig.~\ref{density} for snapshots
of the rest-mass density at select times illustrating aspects of the matter dynamics. 

\section{Results}
\label{results}

Varying the parameters as discussed in the previous section, there is much degeneracy in the qualitative 
features that arise (which will need to be
addressed in future studies investigating extraction of source properties from GW and EM observations).  
This
is essentially because the leading-order source of the variability 
is rooted in the following two properties of the system:
(1) the NS radius, varied by altering the EOS (as the NS mass is fixed in this study), 
(2) the location of the innermost stable orbit (ISO) varied by changing the spin of 
the BH or the eccentricity of the encounter.
For equatorial geodesics on a black hole background, the ISOs
asymptote to circular orbits, though these should
not be confused with the innermost {\em stable} circular orbit
(at $r=6M$ in Schwarzschild), or ISCO. The circular
orbits (in the range $r=3M$ to $6M$ in Schwarzschild) associated with the ISOs with 
nonzero eccentricity are {\em unstable}, and under infinitesimal 
perturbation their noncircular nature is manifest in the form 
of zoom-whirl behavior~\cite{Glampedakis:2002ya}. Specifically, depending upon
the size of the perturbation, the geodesic will undergo a number
of near circular orbits (the ``whirls''), followed either (depending upon the sign of the perturbation)
by a plunge into the BH or 
by a single near elliptical orbit (the ``zoom''), and in the latter case the
motion repeats. Away from the geodesic limit there is still an effective
ISO, where qualitatively similar behavior occurs~\cite{Pretorius:2007jn}, though radiation reaction will 
eventually drive the system to a merger for all bound systems.

In terms of the gravitational dynamics, the closer the periapsis is to the ISO, the more
whirling that occurs, resulting in enhanced GW emission and more rapid evolution
of effective orbital parameters. Regarding the matter dynamics,
the ISO is essentially the ``event horizon'' for fluid elements following 
geodesics, or being close to geodesic. What this implies is on a close-encounter, if the NS is not
disrupted and the periapsis of the orbit is within the ISO a merger will result
and the entire NS will fall into the BH. If the NS is disrupted, following the
essentially Newtonian redistribution of angular momentum that results,
fluid elements within an ISO corresponding to their effective eccentricity
will immediately plunge into the BH, while the rest will either move out onto
eccentric orbits to later fall back onto an accretion disk or be ejected from the system.
If the majority of the NS mass ends up outside the ISO following disruption, once it
moves beyond the tidal-disruption zone there will be sufficient self-gravity for
the material to recoalesce into a NS core. Similarly,
in a partial disruption where only the outer layers are stripped from the
NS, some stripped material will accrete into the BH, some back onto the NS, and a portion will 
be flung out unbound.

In the following subsections, we break down the discussion of phenomenology
of the encounters by first summarizing the results from the base case
presented in~\cite{ebhns_letter} (Sec.~\ref{base_case}), then describe in turn what
happens when spin (Sec.~\ref{sec_spin}), EOS (Sec.~\ref{sec_eos}), and
eccentricity (Sec.~\ref{sec_ecc}) are changed relative to the base case.
In order to have some intuitive understanding
of why the changes have the effects they do,
it is useful to keep the above discussion in mind.  This can also allow one to anticipate
what would happen if parameters we are not altering here are changed, for example,
the BH or NS mass.

\subsection{Zero-spin survey with HB EOS}
\label{base_case}
\begin{figure}
\begin{center}
\includegraphics[width=3.2in,clip=true,draft=false]{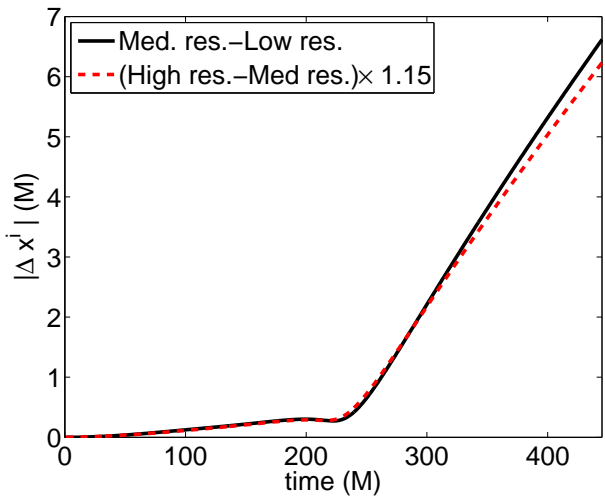}
\end{center}
\caption{
Convergence of the NS trajectory for the $r_p=10, a=0$-case (see Sec.~\ref{base_case}). 
What is plotted is the
magnitude of the difference in the center of rest-mass position of the NS between
the low and mediumi-resolution runs, and between the medium and high-resolution
runs scaled assuming second-order convergence. The close-encounter
occurs near $t=230M$, where the largest ``perturbation'' of the orbit
due to truncation error happens; the relatively large growth of
the difference in trajectories following this is mostly just a reflection
of this change in the orbital parameters that occurred around periapsis.
}
\label{rp10_nscom_conv}
\end{figure}

Here we summarize the results obtained for simulations with an initially 
nonspinning BH and a NS with the HB EOS (see \cite{ebhns_letter} for more details). 
We considered a range 
of periapsis separations from $r_p/M=5.0$ to 15 ({\it i.e.,} 50 to 150~km). 
Henceforth, we will consider $r_p$ to be normalized by $M$.
In all of these cases, sufficient energy is carried
away by GWs to result in a bound system.  Our simulations exhibit three types
of behavior: (1) a direct plunge
($r_p=5.0,5.83,6.67,6.81$); (2) following the initial periapsis
passage, a single elliptical orbit and then a plunge
($r_p=6.95,7.22,7.5$); and (3) following the initial periapsis
passage, a long-period elliptical orbit ($r_p=8.75,10.0,12.5,15.0$).  For the
latter group (and the high-resolution $r_p=7.5$ run), 
we do not simulate the entire orbit since the length of such simulations would make them very computationally expensive
\footnote {For example, based on an estimate using the emitted GWs and assuming a Newtonian orbit the
$r_p=8.75$ case will undergo another close-encounter after $\sim 7000$ M, which would take 2-3
months of wall clock time to simulate at medium resolution (while the larger $r_p$ cases will take even longer).  
Though such long runs are not unheard of, we chose
to use our limited computational resources to explore a greater number of parameters.}, 
and we focus on the burst of GWs associated with the first periapsis passage. 
For one case in each class ($r_p=5.0,7.5,10.0$) we performed a convergence study
which showed approximately second-order convergence and allowed us to perform a Richard 
extrapolation to estimate errors in the resulting GWs.  To give further indication of the truncation error in these runs, in Fig.~\ref{rp10_nscom_conv} we plot the error in the trajectory of the NS in a fly-by case $r_p=10$  
(see Fig.~\ref{trajplot} for the medium resolution trajectory).
The GW energy and angular momentum emitted (including extrapolated values from the resolution studies) 
as well as the disk properties of those cases followed through merger are summarized in Table~\ref{master_table}.  
Table~\ref{merger_bh} shows the spin of the post-merger BH (for this base set of runs
as well as the subsequent parameter survey) for the cases we followed through merger.
As the threshold in $r_p$ dividing (1)
and (2) is approached there is a dramatic enhancement in the gravitational energy and angular momentum 
released during the close-encounter. 

The amount of material remaining after
merger, which could potentially form an sGRB-powering accretion disk also depends 
significantly on impact parameter.  Below the threshold dividing (1) and (2), there is a
sizable amount of remaining material in excess of $20\%$ in the $r_p=6.81$ case. 
In most cases $\approx 50 \%$ of the material is unbound. A simple numerical estimate of the fallback
time for the bound material based on when the elliptic orbit will return to the accretion disk
shows the characteristic $t^{-5/3}$ scaling \cite{rees88}.

\begin{table*}
\begin{center}
{\small
\begin{tabular}{ l l l l l l l l }
\hline\hline
$r_p$      &  $M_0/M_0(t=0)$ $^a$ 
 & $M_{0,u}/M_0(t=0)$ $^b$ 
 & $\tau_{\rm acc}$ (ms) $^c$
 & \multicolumn{2}{c}{First periapsis $^d$}
 & \multicolumn{2}{c}{Total $^e$ } \\
           &                &                   &                      & 
$\frac{E_{GW}}{M}\cdot10^2 $    &  $ \frac{J_{GW}}{M^2}\cdot10^2$ &  $ \frac{E_{GW}}{M}\cdot10^2$  & $\frac{J_{GW}}{M^2}\cdot10^2$  \\
\hline
5.00   &  0.005$^g$   &  0.0     &   25   &   --         &  --         &   $0.67(0.87)^f$ & $4.14(4.86)^f$ \\
6.67   &  0.107   &  0.056   &   130  &   --         &  --         &   $1.29$       & $9.10$ \\
6.81   &  0.221   &  0.101   &   40   &   --         &  --         &   $1.19$       & $9.60$ \\
6.95   &  0.018   &  0.003   &   47   &  $0.697$        &  $7.33$        &   $1.65$       & $13.9$\\
7.22   &  0.013   &  0.001   &   16   &  $0.358$        &  $4.48$        &   $1.18$       & $10.2$ \\
7.50   &  0.009   &  0.003   &   7.6  &  $0.242(0.147)^f$ &  $3.44 (2.46)^f$ &   $1.03$       & $44.7$\\
8.75   &  \ldots  &  \ldots  & \ldots &  $0.073$        &  $1.58$        &   \ldots        & \ldots  \\
10.0   &  \ldots  &  \ldots  & \ldots &  $0.033(0.027)^f$ &  $0.97(0.88)^f$  &   \ldots        & \ldots  \\
12.5   &  \ldots  &  \ldots  & \ldots &  $0.011$        &  $0.46$        &   \ldots        & \ldots  \\
\hline
\end{tabular}
}
\caption{Disk properties and GW energy and angular-momentum losses for an initially
hyperbolic ($e\approx 1$) encounter of a zero-spin BH and NS with HB EOS. Dashed entries
correspond to cases that merge during the first encounter, and hence have no 
``first periapsis''; dotted entries correspond to binaries that were only evolved
through first periapsis passage.
\\
$^a$ Rest-mass remaining outside the
BH shortly ($\sim50$ M) after the end of merger, normalized by the initial total rest-mass.\\
$^b$ Unbound rest-mass estimated using local fluid velocities and assuming a stationary metric.\\
$^c$ Rough {\em initial} accretion timescale ($\tau_{\rm acc} = M_0/\dot{M_0}$) 
evaluated shortly after merger.\\
$^d$ Energy and angular momentum lost to GWs during the first close-encounter.\\
$^e$ Total GW energy and angular-momentum losses for cases which
were followed through merger.\\
$^f$ Results are from medium-resolution runs; values in parentheses are
Richardson-extrapolated estimates using low and high resolutions, where available.
Note that the relatively large error for $r_p=7.5$ (and to a lesser extent
$r_p=5,10$) is due in part
to truncation error altering the actual periapsis by a small
amount, and in this regime the GW emission is highly sensitive to
binary separation (Fig.~\ref{egw_fit}). \\
$^g$ For the $r_p=5$ case $M_0/M_0(t=0)$ was the same at the three resolutions 
to within $\approx 5\times 10^{-4}$; however, for such low disk masses we expect systematic effects,
{\em e.g.} the numerical atmosphere to be important.  \\
}
\label{master_table} 
\end{center}
\end{table*}

\begin{table}
\begin{center}
{\small
\begin{tabular}{ l c c c c c c c c}

\hline\hline
   &  \multicolumn{6}{c}{$a=0$ HB EOS} \\ 
\hline
 $r_p$ & 5.00 & 6.67 & 6.81 & 6.95  & 7.22 & 7.50  \\
 $a_{\rm final}$ & 0.49\footnote{$a_{\rm final}=0.49 \pm 0.01$} & 0.45 & 0.37 & 0.47 & 0.50 & 0.50\\
 $a_{\rm eff}$   & 0.40 & 0.46 & 0.47 & 0.47 & 0.48 & 0.49 \\
\hline
   &  \multicolumn{4}{c|}{$a=0.5$} & \multicolumn{4}{c}{$a=-0.5$} \\ 
\hline
 $r_p$ & 5.00 & 5.50 & 6.00 & \multicolumn{1}{c|}{6.25} & 5.00 & 7.50 & 8.13 & 8.28 \\
 $a_{\rm final}$ & 0.74 & 0.71 & 0.71 & \multicolumn{1}{c|}{0.70} & 0.16 & 0.25 & 0.22 & 0.24\\
 $a_{\rm eff}$   & 0.72 & 0.74 & 0.77 & \multicolumn{1}{c|}{0.77} & 0.08 & 0.17 & 0.19 & 0.19 \\
\hline
   &  \multicolumn{4}{c|}{2H EOS} & \multicolumn{4}{c}{B EOS} \\ 
\hline
 $r_p$ & 5.00 & 6.75 & 7.00 & \multicolumn{1}{c|}{} & 5.00 & 6.25 & 7.00 &  \\
 $a_{\rm final}$ & 0.50 & 0.29 & 0.33 & \multicolumn{1}{c|}{} & 0.48 & 0.52 & 0.48 & \\
 $a_{\rm eff}$   & 0.40 & 0.46 & 0.47 & \multicolumn{1}{c|}{} & 0.40 & 0.45 & 0.47\\

\hline
   &  \multicolumn{2}{c}{$e=0.75$}  \\ 
\hline
 $r_p$ & 7.50 & 7.81 \\
 $a_{\rm final}$ &  0.44 & 0.49 \\
 $a_{\rm eff}$   &  0.46 & 0.47 \\

\hline\hline

\end{tabular}
}
\caption{
Post-merger BH spin (dimensionless) for various initial conditions.  
Also shown is the effective spin $a_{\rm eff}$, used in the model described
in Sec.~\ref{sec_evo_orb} and calculated from
initial conditions using (\ref{a_eff}).}
\label{merger_bh}
\end{center}
\end{table}

\subsection{Effects of black hole spin}\label{sec_spin}
\begin{figure}
\begin{center}
\includegraphics[width=3.2in,clip=true,draft=false]{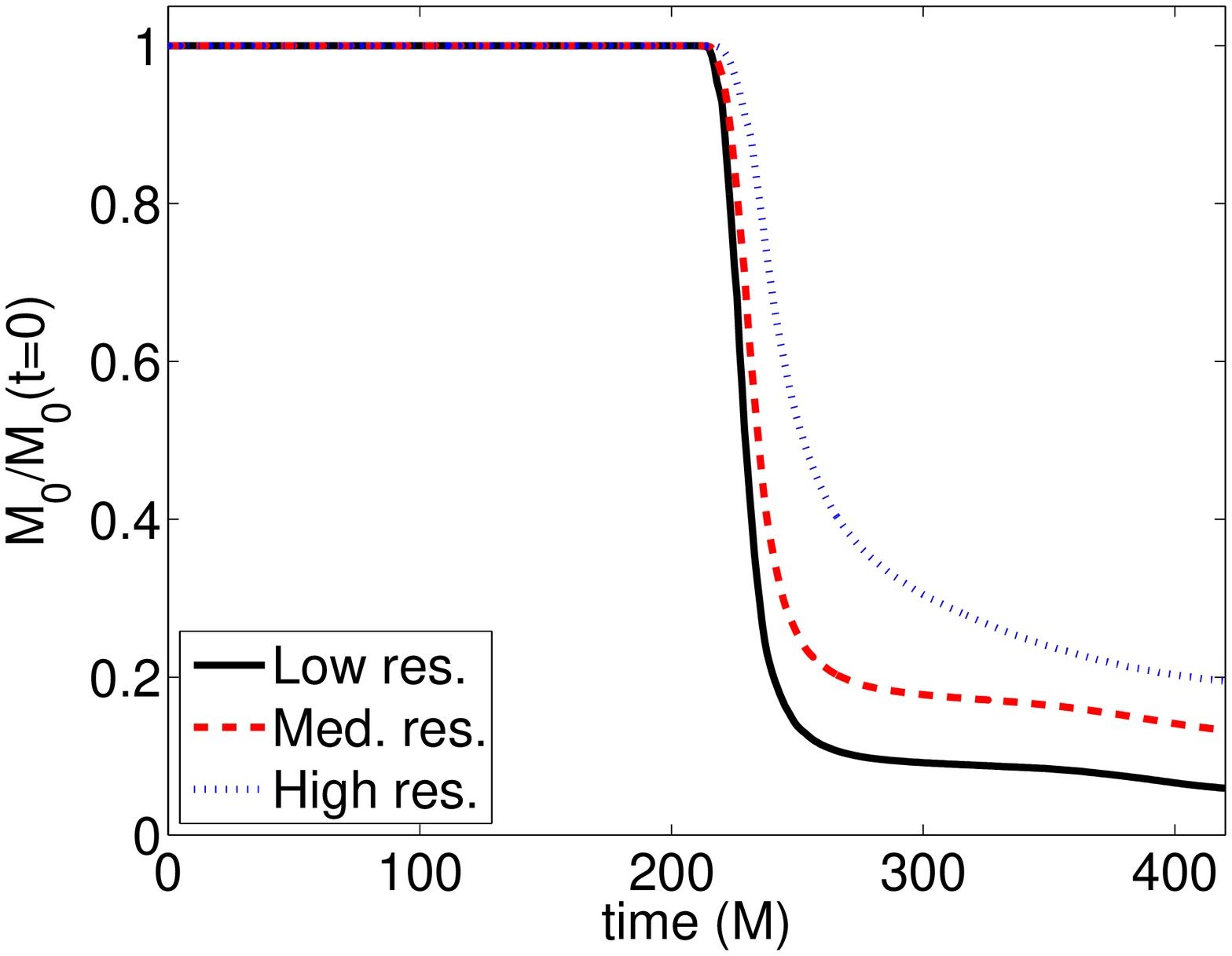}
\end{center}
\caption{
Effect of resolution on the disruption and subsequent accretion
of NS material, as measured by the total rest-mass exterior
to the BH horizon, for the $r_p=5, a=0.5$ case (see Sec.~\ref{sec_spin}). 
The amount of rest-mass remaining at late times at the different resolutions
is consistent with approximately second-order convergence.
}
\label{rp5_ap5_rest_mass}
\end{figure}

As expected from Kerr geodesics, prograde BH spin results in a smaller critical-impact parameter below 
which the NS merges with the BH on the first encounter than in the nonspinning
case; the converse is expected for cases with retrograde initial BH-spin.
For the cases with initial BH-spin $a=0.5$ (top of Table~\ref{spin_table}), only the $r_p=5.0$ case merges 
on the first encounter while $r_p=$ 5.5, 6.0, and
6.25 go back out on a short elliptic orbit before merging.  The $r_p=$ 7.5 and 10.0 cases go out on 
a long elliptic orbit after the initial periapsis passage, and we did not follow these through merger.
For the $r_p=$ 5 case, we also performed convergence tests in order to estimate truncation error 
for a spinning BH where the effective ISO is closer to the BH.  From this we compute Richardson extrapolated
values for the energy and angular momentum of the resulting GW as well as for the amount of 
matter left over post-merger; these values are reported in the table.  In Fig.~\ref{rp5_ap5_rest_mass} we also 
show the amount  of rest-mass exterior to the BH horizon as a function of time and resolution for 
this case.

\begin{figure}
\begin{center}
\includegraphics[width=2.5in,clip=true,draft=false]{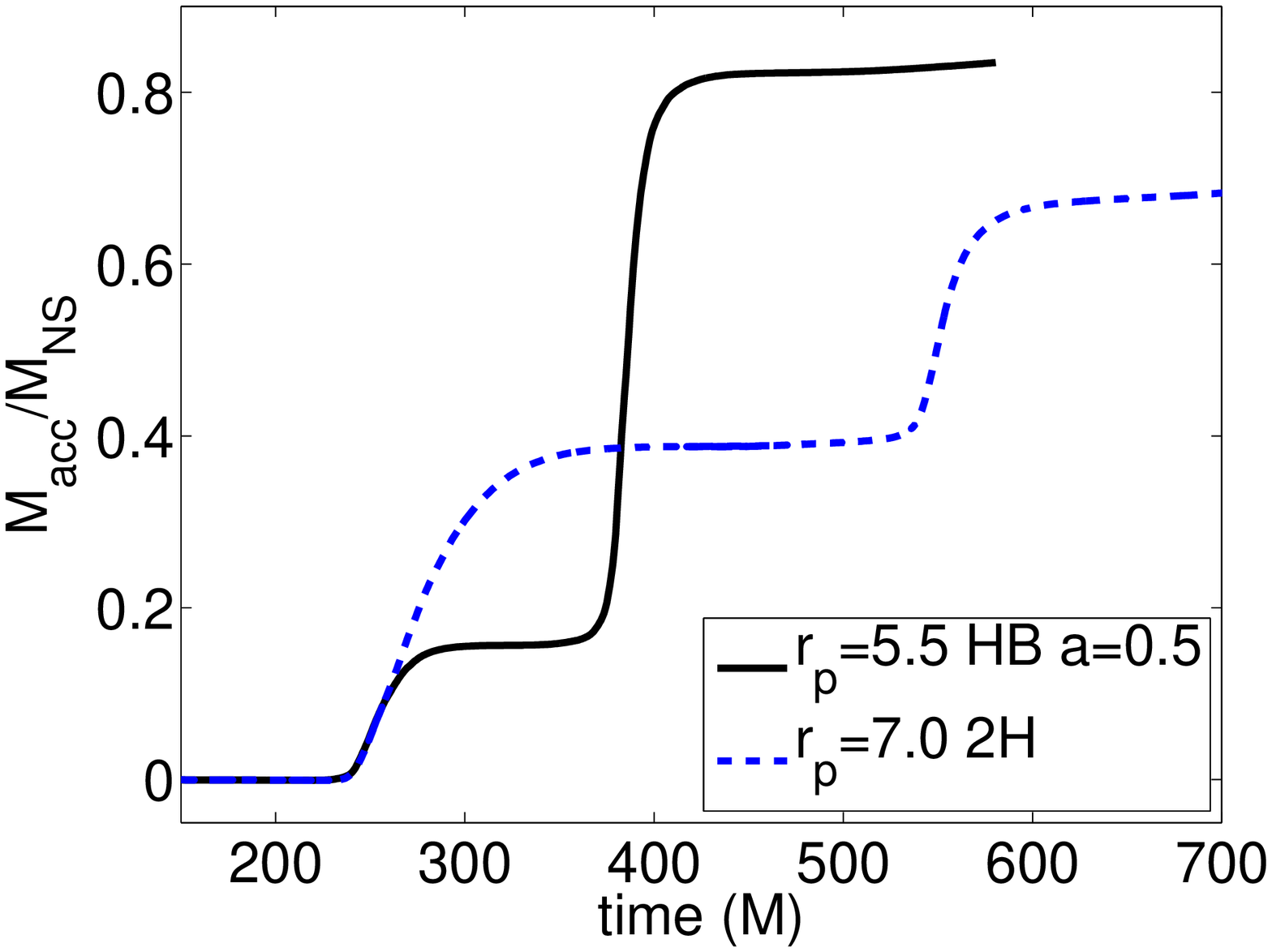}
\includegraphics[width=2.5in,clip=true,draft=false]{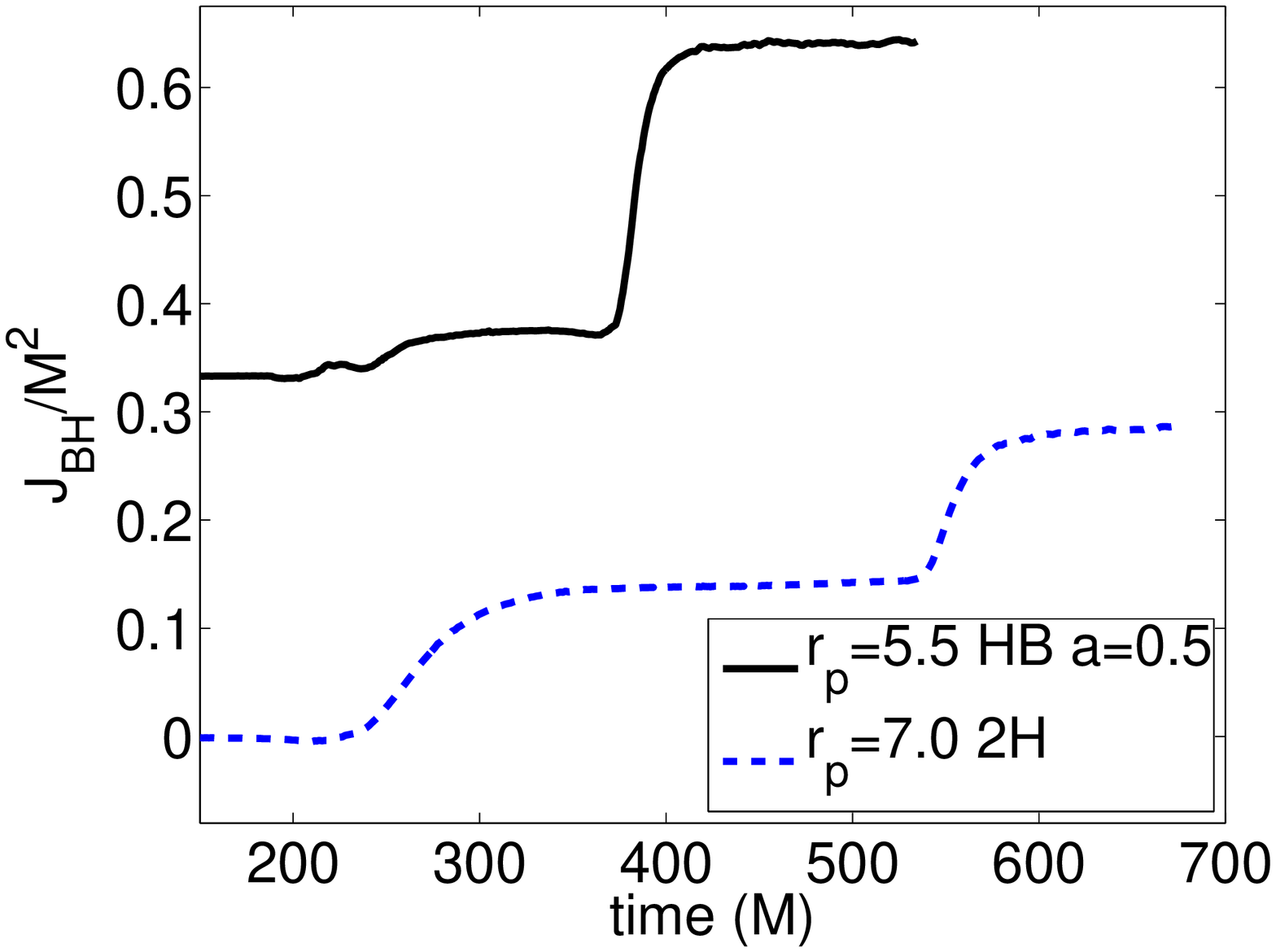}
\caption
{
{\bf Top:} The amount of rest-mass (normalized to the total rest-mass of the NS) accreted by the BH
as a function of time for the $r_p=7.0$ 2H-case and for the $r_p=5.5$, $a=0.5$ HB-case.
In both cases there are two significant episodes of mass transfer.
{\bf Bottom:} The angular momentum of the BH horizon in units of the total ADM-mass squared, $M^2$, as a function of time for the same two cases. 
}
\label{mass_transfer}
\end{center}
\end{figure}

One important consequence of the reduced critical-impact parameter in the prograde case is that around this 
threshold, the tidal forces on the NS are greater and the resulting accretion disks consequently larger.  
Even more striking, the enhanced tidal interaction can lead to significant mass transfer from the NS to the BH 
even in nonmerger close-encounters, as, for example, with the $r_p=5.5$, $a=0.5$ case.  Here the NS becomes highly distorted 
and loses approximately $16\%$
of its mass to the BH (see Fig.~\ref{mass_transfer}).  However, a compact (albeit highly distorted) star remains
(see Fig.~\ref{density} middle, bottom) until it merges with the BH on its second close-encounter.

\begin{figure}
\begin{center}
\includegraphics[width=2.5in,clip=true,draft=false]{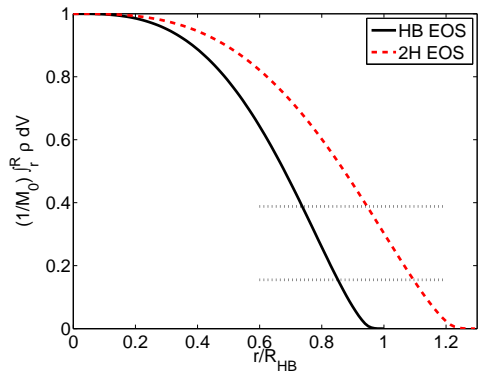}
\caption
{
The fraction of rest-mass outside a given radius $(1/M_0)\int_r^R \rho dV$ 
for an isolated star with the HB or 2H EOS and $M=1.35$ $M_{\odot}$.  Here radius $r$ is normalized to the 
radius of the HB star ($R_{\rm HB}$).  The horizontal, dotted-lines indicate the amount of material
accreted into the BH during the initial close-encounter for the $r_p=5.5$-case with
HB EOS (bottom) and the $r_p=7.0$-case with 2H EOS (top) as shown in Fig.~\ref{mass_transfer}. 
}
\label{mass_profile}
\end{center}
\end{figure}

\begin{figure*}
\begin{center}
\includegraphics[width=3.5in,clip=true,draft=false]{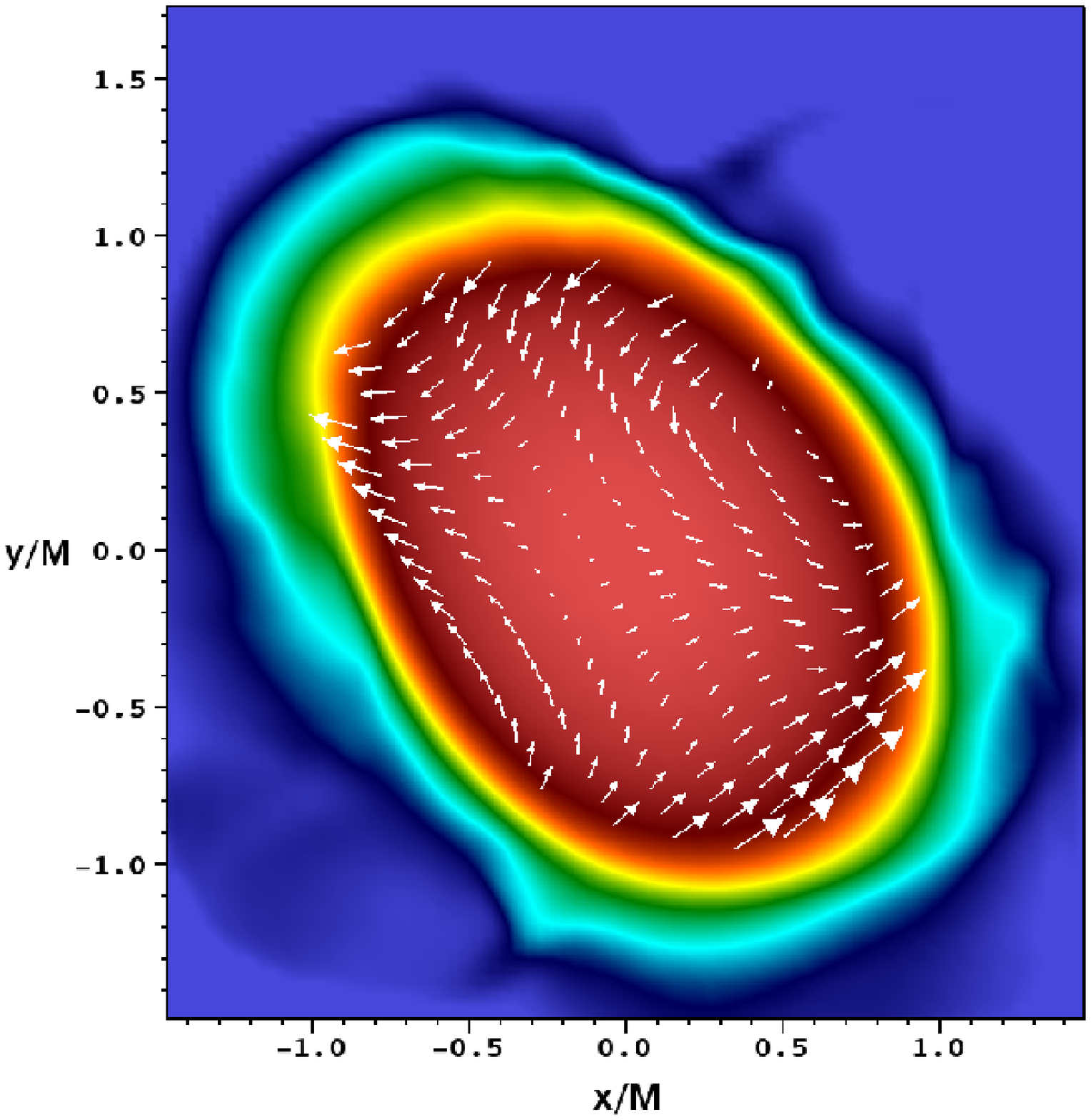}
\includegraphics[width=3.5in,clip=true,draft=false]{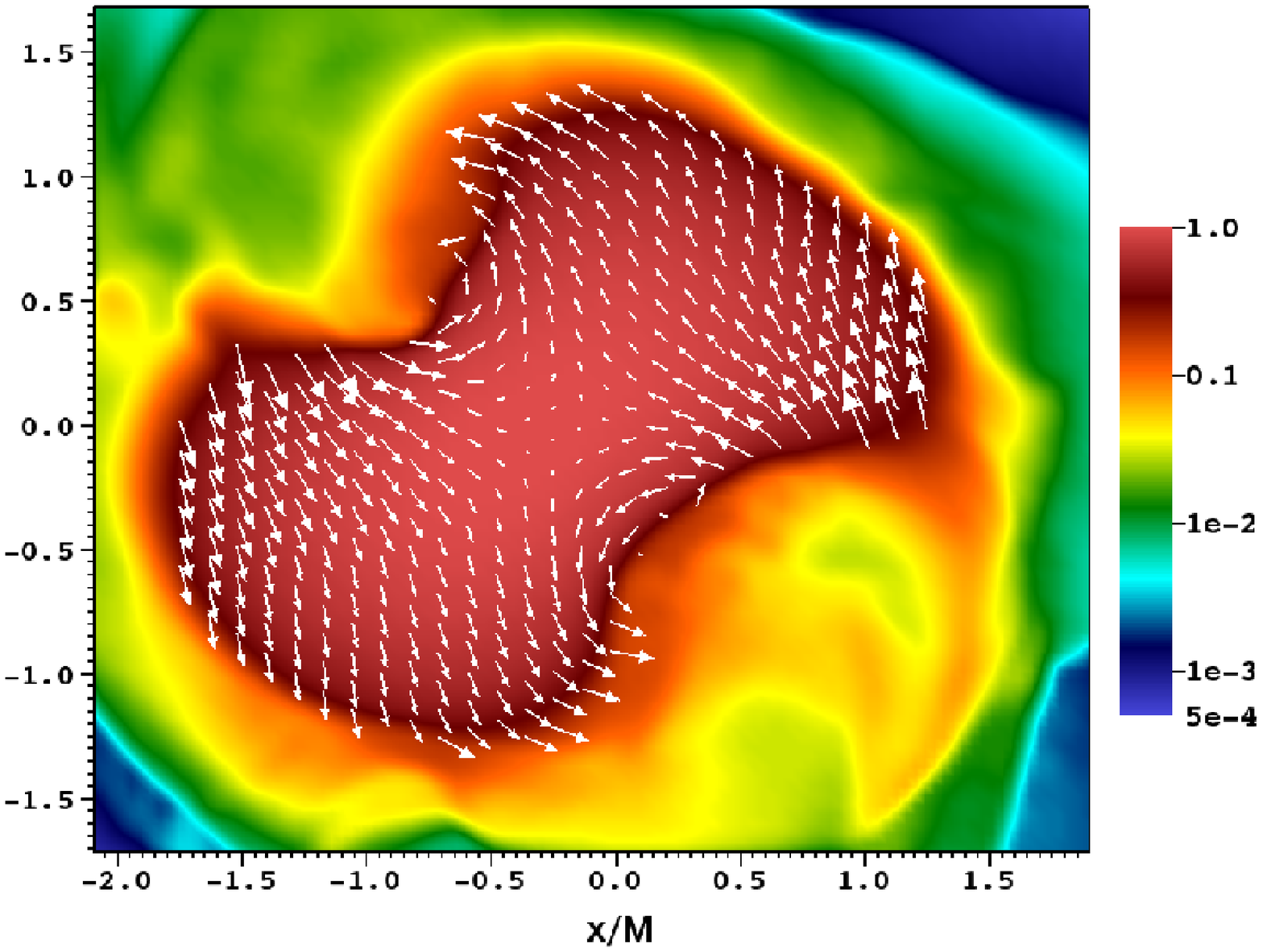}
\caption
{
NSs exhibiting large-amplitude $f$-mode oscillations following the first
interaction with the (nonspinning) BH.  The cases shown are $r_p=6.95$
with the HB EOS at $t=341 M$ (left) and $r_p=7.0$ with the 2H EOS 
at $t=356 M$ (for the latter case, see the bottom left-panel of 
Fig.~\ref{density} for a snapshot of the density near the time of closest approach to the BH
when this large oscillation is excited). The color map shows the rest-mass density in the equatorial
plane on a logarithmic scale, normalized to the instantaneous maximum
density.  The arrows show the velocity in the NS
center-of-mass frame.  The longest velocity arrows correspond to 
velocity magnitudes 0.084 (left) and 0.17 (right).  Though the two 
cases have similar periapsis radii, the lower compaction in the 2H-case
leads to a much stronger tidal interaction and an $f$-mode, which shows
nonlinear characteristics. 
}
\label{f_mode}
\end{center}
\end{figure*}

For the $a=-0.5$ cases (see bottom of Table~\ref{spin_table}), $r_p=$ 5.0, 7.5, and 8.13 are direct mergers
while the $r_p=$ 8.28 case goes through a single elliptic orbit before merging.
The $r_p=$ 8.44, 8.75, and 10.0 cases go out on longer elliptic orbits which we did not follow to completion.  
Because of the larger critical-impact parameter for merger on the first
encounter, there is less tidal disruption and, hence, the accretion disks 
contain $\leq 1\%$ of the total NS rest-mass for all the cases followed through merger.  However,
this also means that the BH and NS can undergo more whirling behavior in the 
critical regime before merger.
This is especially evident in the waveform for the nearer threshold merger case of $r_p=8.13, a=-0.5$ as shown in 
Fig.~\ref{merger_psi4}.  The gravitational wave signal shows several cycles of almost constant 
amplitude and frequency, indicative of a nearly circular orbit.
Compared to the near-threshold nonspinning case ($r_p=6.81$, also shown in Fig.~\ref{merger_psi4}), this whirling
period is much more pronounced in the waveform.

\begin{table*}
\begin{center}
{\small
\begin{tabular}{ l l l l l l l l }
\hline\hline
$r_p$      &  $M_0/M_0(t=0)^a$
 & $M_{0,u}/M_0(t=0)^b$
 & $\tau_{\rm acc}$ (ms) $^c$ 
 & \multicolumn{2}{c}{First periapsis $^d$}
 & \multicolumn{2}{c}{Total$^e$}\\
           &                &                   &                      & 
$\frac{E_{GW}}{M}\cdot10^2 $    &  $ \frac{J_{GW}}{M^2}\cdot10^2$ &  $ \frac{E_{GW}}{M}\cdot10^2$  & $\frac{J_{GW}}{M^2}\cdot10^2$  \\
\hline
$a=+0.5$ \\ 
\hline
5.00   &  $0.165(0.295)^f$ &   $0.021(0.107)^g$ &    4.8 &  --      &  --       &   $ 1.25(1.41)^f$  &  $7.14(9.04)^f$  \\ 
5.50   &  0.174   &   0.088  &    40  &  $0.800$ &  $7.14$   &   $ 1.40$    &  $10.9$    \\  
6.00   &  0.181   &   0.029  &    26  &  $0.360$ &  $4.16$   &   $1.53$     &  $14.7$    \\ 
6.25   &  0.080   &   0.014  &    33  &  $0.347$ &  $4.11$   &   $1.29$     & $12.0$   \\  
7.50   &  \ldots  &  \ldots  & \ldots &  $0.104$ &  $1.90$   &   \ldots     & \ldots     \\
10.0   &  \ldots  &  \ldots  & \ldots &  $0.025$ &  $0.82$   &   \ldots     & \ldots  \\

\hline
$a=-0.5$ \\ 
\hline
5.00   &   0.007  &    0.0   &  36    &  --             &  --            &   $ 0.33$       &  $2.32$  \\ 
7.50   &   0.008  &    0.0   &  71    &  --             &  --            &   $0.82$        & $5.91$\\
8.13   &   0.010  &    0.0   &  0.12  &  --             &  --            &    $1.57$       &   $13.7$ \\
8.28   &   0.002  &    0.0   &  2.8   &  $0.385$        &  $5.12$        &   $1.25$        & $12.8$ \\ 
8.44   &  \ldots  &  \ldots  & \ldots &  $0.268$        &  $3.98$        &   \ldots        & \ldots  \\
8.75   &  \ldots  &  \ldots  & \ldots &  $0.167$        &  $2.85$        &   \ldots        & \ldots  \\
10.0   &  \ldots  &  \ldots  & \ldots &  $0.052$        &  $1.32$        &   \ldots        & \ldots  \\
\end{tabular}
}
\caption{
Disk properties and GW energy and angular-momentum losses for an initially
hyperbolic ($e\approx 1$) encounter of a BH with spin $a=\pm0.5$ and NS with HB EOS. The same
comments and set of footnotes $^a$ to $^f$ apply as in Table~\ref{master_table}. 
\\
$^g$ For the $r_p=5$, $a=+0.5$ case the Richardson extrapolated value of $M_{0,u}/M_0(t=0)$ was 
computed using just the medium and high-resolution results and assuming second-order convergence
due to the low amount of unbound material in the low-resolution case. \\
}
\label{spin_table} 
\end{center}
\end{table*}

\subsection{Effects of equation of state}\label{sec_eos}

We also consider the B EOS, which is softer than the HB, and the 2H EOS, which is stiffer.  
Though the different EOSs have only a slight effect on the critical impact parameter, they can
have a dramatic effect on the near-threshold behavior.  For the B EOS, we chose four different
impact parameters (bottom of Table~\ref{eos_table}): $r_p=$ 5.0 and 6.25 (direct mergers); $r_p=$ 7.0 (after periapsis passage, a short elliptic orbit and then merger);
and $r_p=$ 7.5 which we did not follow through its full elliptic orbit after the first close-encounter. 
The more compact B EOS NS experiences less
tidal disruption and forms accretion disks with $\leq 1\%$ of the total NS rest-mass for all cases
followed through merger.  
For the 2H EOS, we considered five different 
impact parameters (see top of Table~\ref{eos_table}): direct mergers $r_p=$ 5.0 and 6.75; $r_p=$ 7.0 (after periapsis passage, a short elliptic orbit and then merger);
and $r_p=$ 7.25 and 7.5 which we did not follow through the full elliptic orbit after the first close-encounter. 

The less compact 2H EOS NS experiences significant tidal deformation, and for the $r_p=$ 6.75 and 7.0 cases
there is $\approx 30\%$ of the NS material leftover post-merger.
In the case of $r_p=7.0$ the deformation and tidally-induced oscillations of the NS are especially severe 
(see Fig.~\ref{density}, bottom left) and there is 
a period of significant mass transfer (Fig.~\ref{mass_transfer}) from the NS to the BH during the
first close-encounter.  This is very similar to the $r_p=5.5, a=0.5$ case, though more pronounced.
For both the 2H EOS and the $a=0.5$ with HB EOS simulations no significant mass transfer
is found for runs with $r_p$ slightly above 7.0 and 5.5, respectively, so we can use these as an effective
measure of the separation $d_{R}$ at which the star begins to overflow its Roche lobe.  Though the individual
values are somewhat higher than expected from Newtonian theory, it predicts
a scaling with compaction $d_{R}\propto \mathcal{C}_{NS}^{-1}$, where
$\mathcal{C}_{NS}\equiv M_{NS}/R_{NS}$; this does approximately hold here since $C_{HB}/C_{2H}=1.3$.
This ignores any relativistic effects, both gauge and physical, the latter including, for 
example, the effect of BH spin 
on tidal disruption. In Fig.~\ref{mass_profile} we show the mass profiles of an isolated star 
with the 2H and HB EOS.  This shows that the outer spherical shell that contains
the approximately $40\%$ of the NS material that is accreted into the BH during the first close-encounter for 2H
corresponds to a volume containing almost no material in the HB case.  Again, though
we are ignoring the complicated details of mass-transfer dynamics, this is suggestive as to why, for the 
BH-NS system studied here, there is significant mass transfer for the 2H EOS around $r_p=7.0$
but very little for the HB EOS. 

Since for the 2H $r_p=7.0$ case so much mass is transferred before the merger 
and there is a strong disruption during merger,
the gravitational wave signal resulting from merger itself is significantly weaker than for other cases 
(see Fig.~\ref{merger_psi4}).  The full waveform from Fig.~\ref{rp7_2H_psi4} also shows that 
the GW pulse from the fly-by
dominates the signal compared to the merger part of the waveform.  

Between the initial close-encounter and merger there is also evidence in the GW signal of excited $f$-modes within the NS;
this was also observed in eccentric-binary NS encounters~\cite{2011arXiv1109.5128G}. 
Here we give a qualitative account of the stellar dynamics to show
that the dominant oscillation is akin to that of an $f$-mode of an isolated, perturbed
star---a detailed study, aside from the difficulty of applying a perturbative analysis
in such a transient and in some cases highly distorted star, is beyond the scope of this study.
Figure~\ref{f_mode} demonstrates the density and velocity distortions
for the $r_p=7.0$, 2H case (right panel) as well as the $r_p=6.95$ HB case in which
the distortion is less extreme (left panel).  The latter bears a
particularly strong resemblance to the pure $l=m=2$
$f$-mode flow pattern (see, {\em e.g.}, Fig.~19 of \cite{kastaun}).  In both cases, an animation
of the density field seems to suggest a rotating, distorted NS.  However, as Fig.~\ref{f_mode}
demonstrates, the velocity pattern is not one of overall rotation but rather that of an
oscillatory mode.  Indeed, the circulation theorem should hold to a good approximation in the
bulk of the NS material (though there are entropy-generating shocks near the surface).  Thus, the
tidal interaction with the BH would not induce rotation in the usual sense, but rather oscillatory
modes with rotating patterns.  Many such modes are likely to be excited, but the
$l=m=2$ $f$-mode seems to dominate.  We performed a spherical-harmonic decomposition of the star's 
rest-mass density $\rho = \sum C_{lm}Y_{lm}$ at select times and found $C_{22}$ to be the 
largest coefficient next to $C_{00}$.
We also checked that the perturbation amplitude grows monotonically with radius.
For the $r_p=6.95$ simulation with the HB EOS $|C_{22}/C_{00}| \approx 0.01-0.02$ at
$r=0.4M$ grows to $|C_{22}/C_{00}|\approx0.15-0.20$ at $r=0.8M$. Here $r$ is the radius of the sphere centered on the NS center-of-mass on which the coefficients are calculated. 
The next largest coefficient $C_{20}$ is smaller by a factor of $\gtrsim 2$.  The $r_p=7.0$ simulation with the 2H EOS shows similar behavior although the coefficients are somewhat larger
with $|C_{22}/C_{00}|\approx0.03$ and 0.25 at $r=0.4M$ and $0.8M$, respectively. 

We briefly comment on the possible detectability of such an $f$-mode excitation
in GWs. Such an observation in principle could provide a wealth of information
about the structure of the NS, in particular, since $f$-mode frequencies are
quite sensitive to the EOS (see for example~\cite{Kokkotas:1999mn}). However,
here, (a) the amplitudes are quite low relative to the dominant GW emission
(see Fig.~\ref{rp7_2H_psi4}), (b)
for the cases where the largest amplitudes are excited, the initial $r_p$ is sufficiently
small that only a few cycles of waves will be emitted before subsequent
merger, limiting the signal-to-noise that could be built up, and (c) the frequency
is quite high (above 1 kHz) and thus not in a regime where the AdLIGO-class
detectors are very sensitive. Thus, even if there is a sizable population of
eccentric merger events as studied here, it is unlikely that any corresponding
$f$-mode excitation will be observed with the current generation of GW detectors.

\begin{table*}
\begin{center}
{\small
\begin{tabular}{ l l l l l l l l }
\hline\hline
$r_p$      &  $M_0/M_0(t=0)^a$
 & $M_{0,u}/M_0(t=0)^b$
 & $\tau_{\rm acc}$ (ms) $^c$
 & \multicolumn{2}{c}{First periapsis $^d$}
 & \multicolumn{2}{c}{Total $^e$} \\
           &                &                   &                      & 
$\frac{E_{GW}}{M}\cdot10^2 $    &  $ \frac{J_{GW}}{M^2}\cdot10^2$ &  $ \frac{E_{GW}}{M}\cdot10^2$  & $\frac{J_{GW}}{M^2}\cdot10^2$  \\
\hline
2H EOS \\ 
\hline
5.00   &   0.008  &   0.0    &   56    &  --             &  --        &   $0.58$       &    $3.61$   \\ 
6.75   &   0.278  &   0.117  &   18    &  --             &  --        &   $0.47$       &    $4.65$   \\ 
7.00   &   0.303  &   0.149  &   60    &  $0.387$        &  $4.53$    &   $0.43$       &    $5.41$   \\  
7.25   &   \ldots & \ldots   &  \ldots &  $0.283$        &  $3.90$    &    \ldots      &     \ldots      \\
7.50   &   \ldots &  \ldots  &  \ldots &  $0.200$        &  $3.07$    &   \ldots       &    \ldots   \\

\hline
B EOS \\ 
\hline
5.00   &   0.008  &    0.0   &    57  &  --             &  --        &   $0.60$        &  $3.83$  \\  
6.25   &   0.008  &    0.0   &    64  &  --             &  --        &   $0.87$        &  $5.63$  \\
7.00   &   0.010  &    0.001 &    10  &  $0.718$        &  $7.69$        &   $1.73$        &  $10.7$  \\
7.50   &  \ldots  &  \ldots  & \ldots &  $0.268$        &  $3.72$        &   \ldots        & \ldots   \\

\end{tabular}
}
\caption
{
Disk properties and GW energy and angular-momentum losses for an initially
hyperbolic ($e\approx 1$) encounter of a BH with zero-spin and a NS with 2H and B EOS. The same
comments and set of footnotes $^a$ to $^f$ apply as in Table~\ref{master_table}. 
}
\label{eos_table} 
\end{center}
\end{table*}

\subsection{Bound eccentric evolution}\label{sec_ecc}
\begin{figure}
\begin{center}
\includegraphics[width=2.5in,clip=true,draft=false]{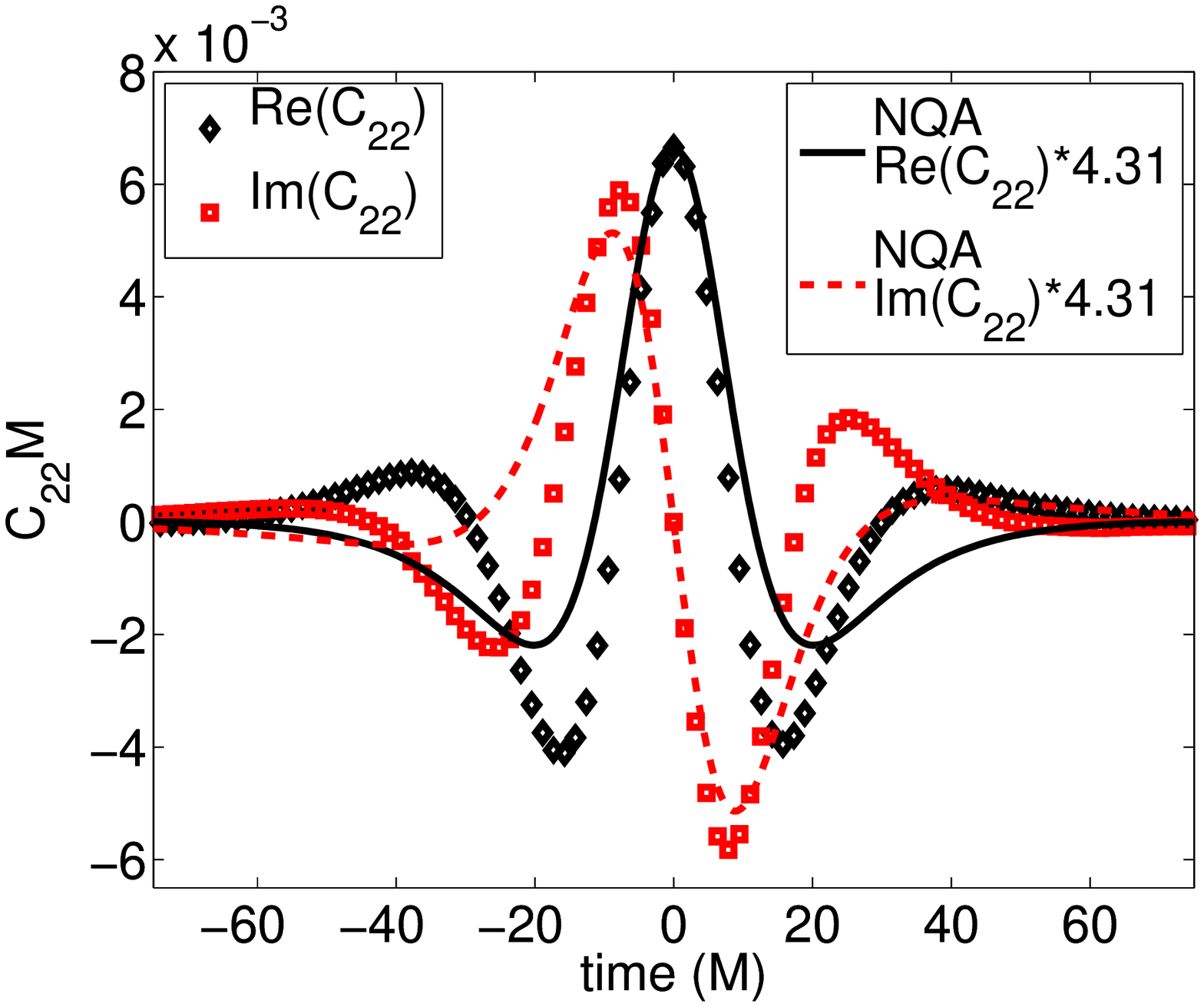}
\includegraphics[width=2.5in,clip=true,draft=false]{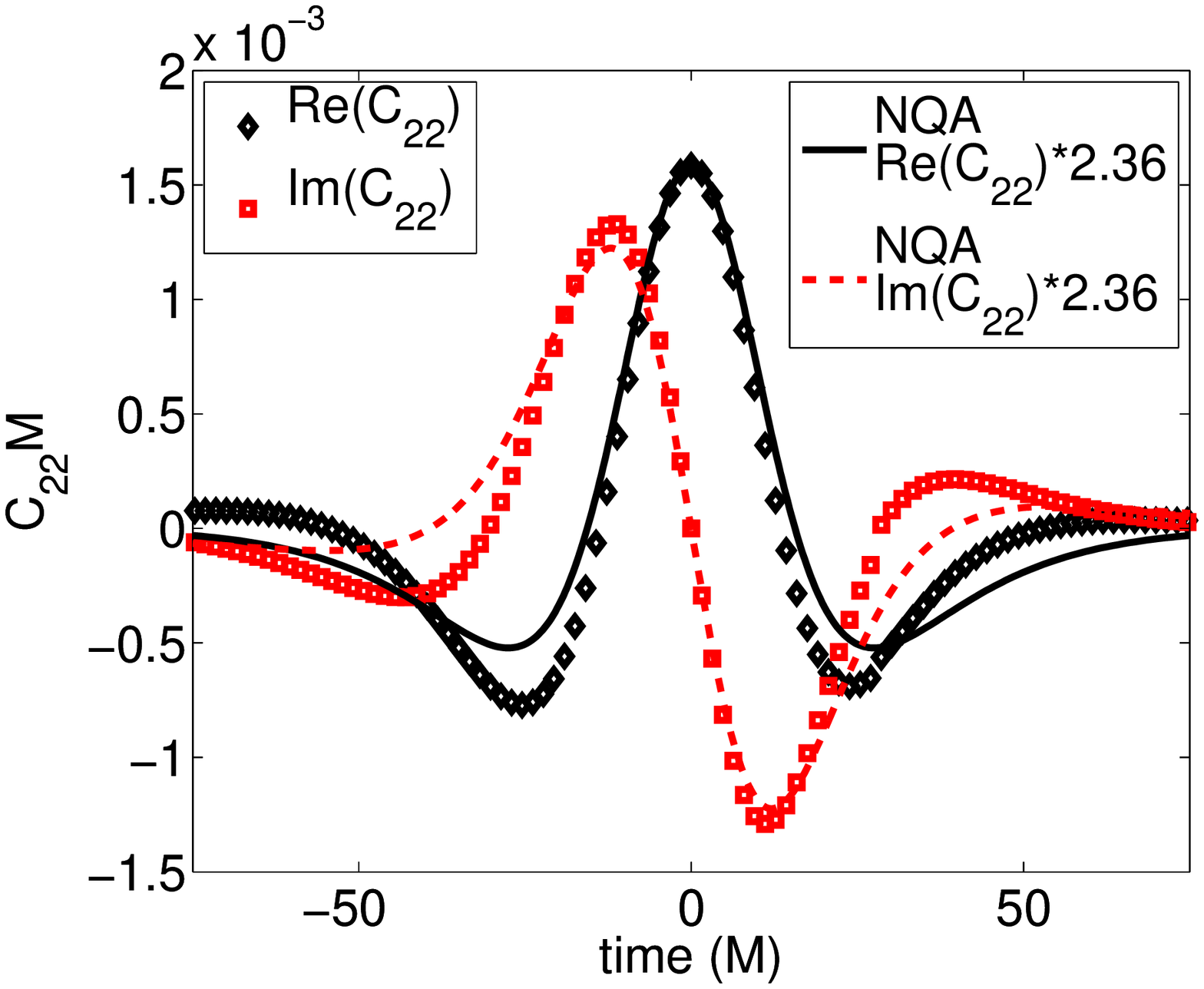}
\caption
{
The real and imaginary components (black diamonds and red squares) of the $l=2$, $m=2$ 
spherical-harmonic of $r\Psi_4$ for $e=0.75$ and $r_p=8.125$ (top) and $r_p=10$ 
(bottom).  For comparison the NQA analytical 
results are shown multiplied by an overall factor so that the magnitude and 
phase match at peak ($t=0$).
}
\label{compare_turner}
\end{center}
\end{figure}

\begin{figure}
\begin{center}
\includegraphics[width=3.4in,clip=true,draft=false]{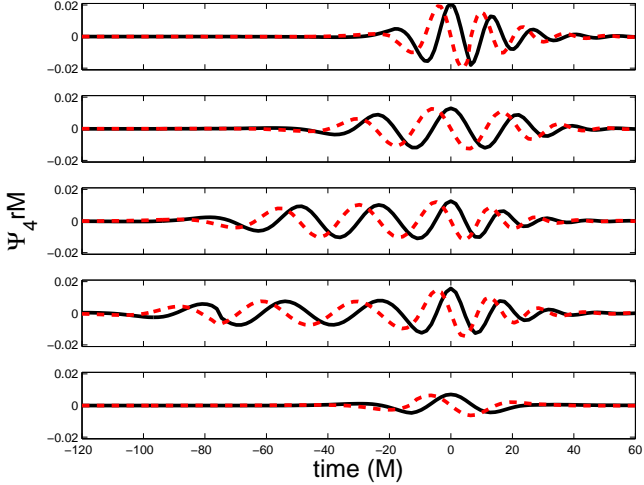}
\caption
{
The real and imaginary components (solid, black-lines  and red, dotted-lines) of 
$\Psi_4$ on the z-axis (perpendicular to the orbital plane) during merger
for the following cases (top to bottom): $r_p=5$, HB; $r_p=6.81$, HB; $r_p=7.5, e=0.75$, HB;
$r_p=8.13, a=-0.5$, HB; and $r_p=7.0$, 2H (see Fig.~\ref{rp7_2H_psi4} for the full signal).
The waveforms are aligned so that the peak occurs at $t=0$ with zero phase.
}
\label{merger_psi4}
\end{center}
\end{figure}

\begin{figure}
\begin{center}
\includegraphics[width=3in,clip=true,draft=false]{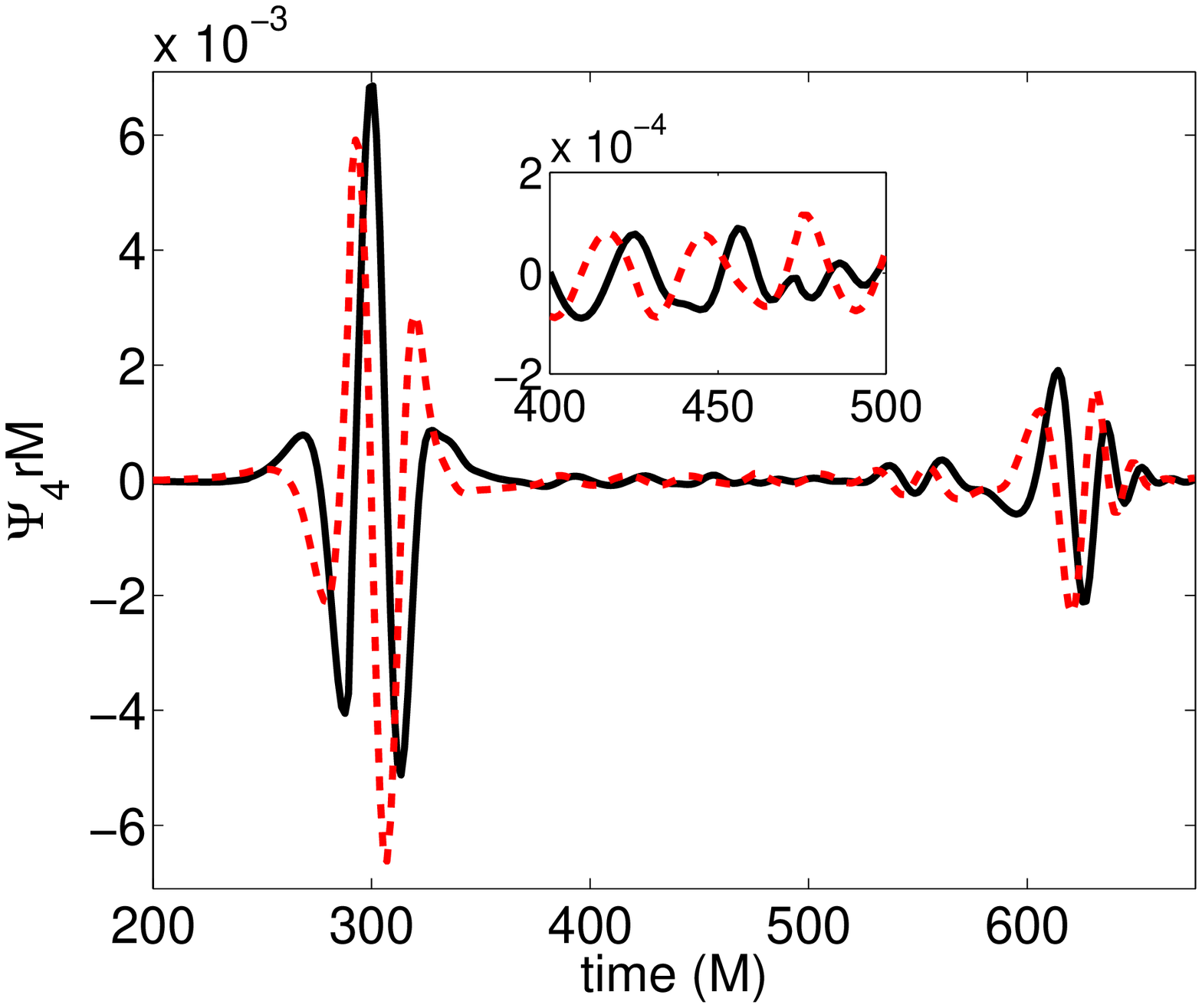}
\caption
{
The real and imaginary components (solid black lines  and red dotted lines) of 
$\Psi_4$ on the z-axis (perpendicular to the orbital plane) at $r=70$ M for the $r_p=7.0$ 2H simulation.
The large burst at $t\approx300$ M comes from the initial fly-by, where the NS becomes extremely distorted 
(Fig.~\ref{density}) and loses a significant portion of its mass (Fig.~\ref{mass_transfer}).  The smaller pulse at
$t=600$ M comes from the merger.  In between, there is a smaller component of the signal coming from the 
tidally induced oscillation of the NS.
}
\label{rp7_2H_psi4}
\end{center}
\end{figure}

Finally, we simulated binaries where the BH and NS have initial-orbital
parameters corresponding to a bound orbit with $e=0.75$ for a range of values of $r_p$.  
These simulations can be viewed as corresponding 
to systems that have already become bound through one or more close-encounters and lost
some of their initial eccentricity.  In 
practice, the length of such evolutions would be very computationally expensive to follow in full.
Of the impact parameters considered (see Table~\ref{bound_table}),
$r_p=$ 7.5 is the only direct-plunge, and $r_p=$ 7.81 is the only case we followed past periapsis passage
to merge on the second close-encounter.  For the remaining cases ($r_p=$ 8.13, 8.75, and 10.0), we only followed
partially through their elliptic orbits after the first close-encounter. 
The threshold for merger on the first encounter moves out slightly in this case
as an appeal to geodesics in Schwarzschild would suggest.
In Fig.~\ref{compare_turner}, we plot the gravitational waveforms from two fly-by close-encounters $r_p=$ 8.13 and 10.0.  In addition, we fit the expected waveform
according to a Newtonian-order quadrupole approximation~\cite{pm63,turner77,Berry:2010gt} (NQA) to our numerical results by
multiplying by an overall factor.  As found in~\cite{ebhns_letter} and in our other simulations here 
for the marginally unbound case, though the shape 
of the waveforms agree quite well, the numerical results exhibit a significant amplitude enhancement that is larger the closer
one gets to the threshold for merger during the close-encounter.
In addition, from Fig.~\ref{merger_psi4} we can see that the near-threshold merger waveform shows more evidence of whirling than the initially unbound case.
  
\begin{table*}
\begin{center}
{\small
\begin{tabular}{ l l l l l l l l }
\hline\hline
$r_p$      &  $M_0/M_0(t=0)^a$
 & $M_{0,u}/M_0(t=0)^b$
 & $\tau_{\rm acc}$ (ms) $^c$
 & \multicolumn{2}{c}{First periapsis $^d$}
 & \multicolumn{2}{c}{Total $^e$}\\
           &                &                   &                      & 
$\frac{E_{GW}}{M}\cdot10^2 $    &  $ \frac{J_{GW}}{M^2}\cdot10^2$ &  $ \frac{E_{GW}}{M}\cdot10^2$  & $\frac{J_{GW}}{M^2}\cdot10^2$  \\
\hline
7.50   &   0.109  &   0.062  &   32   &  --             &  --            &   $1.47$        & $13.6$ \\ 
7.81   &   0.007  &   0.001  &   7.1  &  $0.410$        &  $5.25$        &   $1.58$        & $16.1$  \\ 
8.13   &  \ldots  &  \ldots  & \ldots &  $0.248$        &  $3.67$        &   \ldots        & \ldots  \\
8.75   &  \ldots  &  \ldots  & \ldots &  $0.125$        &  $2.34$        &   \ldots        & \ldots  \\
10.0   &  \ldots  &  \ldots  & \ldots &  $0.049$        &  $1.28$        &   \ldots        & \ldots  \\
\end{tabular}
}
\caption
{
Disk properties and GW energy and angular-momentum losses for an initially
eccentric $e=0.75$ encounter of a zero-spin BH with a NS with HB EOS. The same
comments and set of footnotes $^a$ to $^f$ apply as in Table~\ref{master_table}. 
}

\label{bound_table} 

\end{center}
\end{table*}

\section{Evolution of orbital parameters}\label{sec_evo_orb}

As noted in \cite{ebhns_letter}, there is a dramatic enhancement in 
gravitational-wave energy and angular momentum emitted in a BH-NS close-encounter that occurs 
as the threshold for merger on the first encounter 
is approached.  In this section, we will attempt to explain this enhancement
by analogy to the zoom-whirl orbits in Kerr spacetimes,
and extrapolate this behavior to close-encounters with different eccentricities
and spins.  Then we will use these results to approximate the evolution of
orbital parameters for BH-NS systems undergoing a series of close-encounters.

\subsection{Zoom-whirl enhancement}
\begin{figure}
\begin{center}
\includegraphics[width=2.5in,clip=true,draft=false]{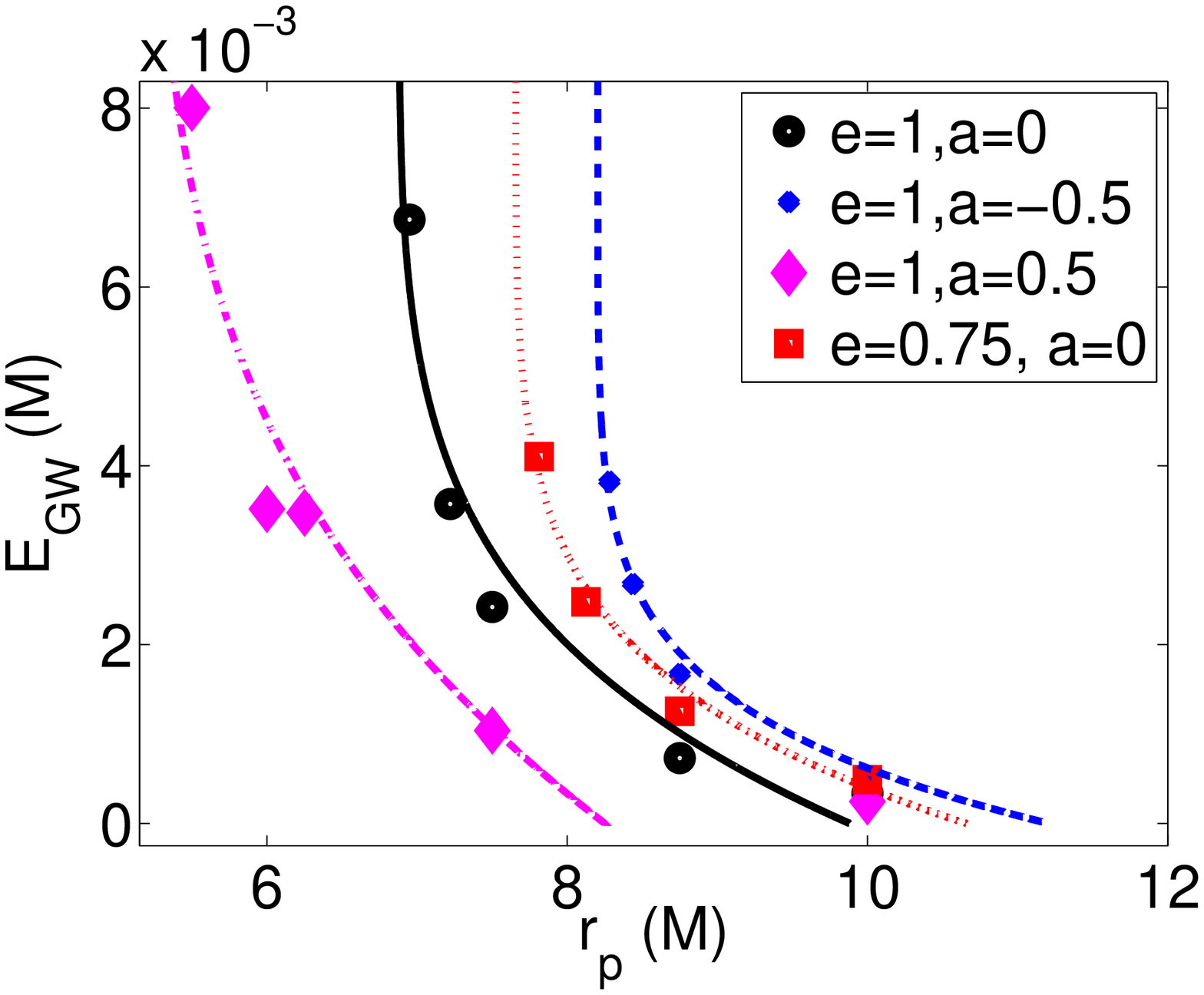}
\includegraphics[width=2.5in,clip=true,draft=false]{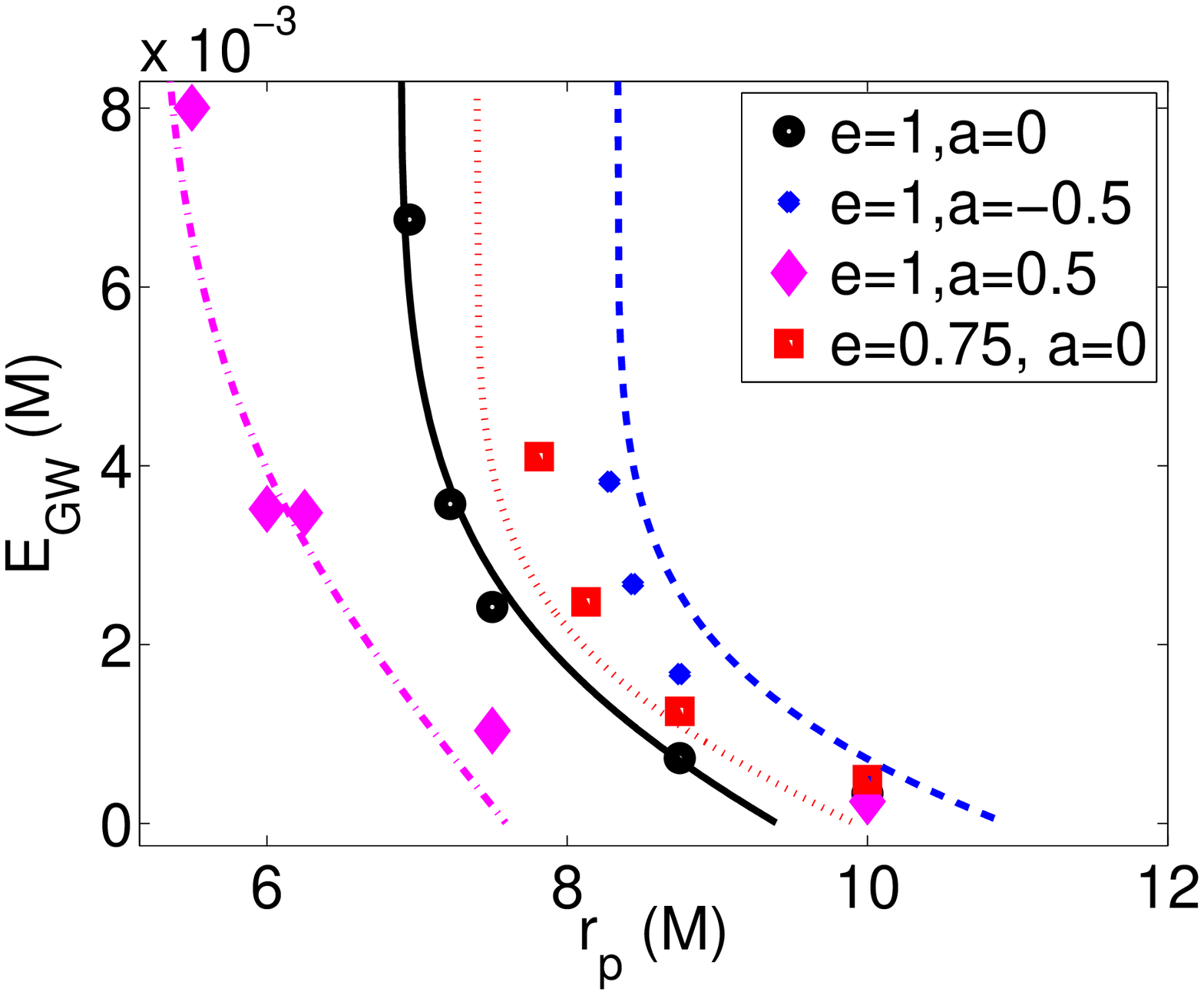}
\caption
{
{\bf Top:} Energy lost to GWs during the initial close-encounter (i.e. excluding merger) as a function of $r_p$
for initial BH spin $a=0$,$-0.5$, and $0.5$ and for initial eccentricity $e=1$ and 0.75.
The functional form $E_0(1-(\delta r_p/\Delta)^\gamma)$ (lines) motivated
by zoom-whirl dynamics is a fit
to the simulation results (points).
$\delta r_p =r_p-r_c$ where $r_c$ is the threshold value
for merger during the encounter.
$E_0$ is the difference in energy between a 
quasicircular orbit and an $e\approx 1$ (0.75) orbit, both having 
$r_p=r_c$.  $\Delta$ is the range over which zoom-whirl-like behavior
dominates the GW-emission energetics. $\gamma$ is a parameter
that in the geodesic analogue is related to the
instability exponent of the corresponding unstable circular
orbit, here, we use it as our fitting parameter.
We obtain $\gamma=0.19$, 0.13, 0.25, and 0.16  for 
$(e,a)=(1,0)$, (1,-0.5), (1,0.5), and (0.75,0), respectively.
{\bf Bottom:} This shows the same simulation data points as the top figure,
though here we use the $e=1$, $a=0$ case (solid, black points) to determine
the free parameters for the method described in the text 
to extrapolate the values of $r_c$ and $\gamma$ to the other three cases.
}
\label{egw_fit}
\end{center}
\end{figure}

As discussed in the previous section and as can be seen in Fig.~\ref{egw_fit}, though different 
values of initial spin or eccentricity may shift the critical value of $r_p$, which we shall call $r_c$ 
at which the NS merges
with the BH during a first encounter, for all the cases considered there is
a significant enhancement in the energy of the gravitational waves emitted
as this threshold is approached.  This can be understood by analogy to the zoom-whirl orbits
in Kerr spacetimes. As the BH-NS approaches the critical-impact parameter,
it is closer and closer to the unstable orbit dividing plunging and nonplunging 
orbits and will, therefore, exhibit more and more whirling behavior.  Though the effect is
most dramatic near threshold, this zoom-whirl analog is a way to understand the significant enhancements
over the NQA predictions that persist till $r_p\approx 10$ (see Fig.4 in~\cite{ebhns_letter}).
This motivates a fit to the GW energy emitted in a whirling close-encounter
with the functional form (see Fig.~\ref{egw_fit})
\begin{equation}
E_{\rm GW}=E_0(1-(\delta r_p/\Delta)^\gamma)
\end{equation}
where $\delta r_p=r_p-r_c$,
$E_0$ is the difference in energy between a 
quasicircular orbit and an eccentricity $e$ orbit both with 
$r_p=r_c$, $\Delta$ is the range over which zoom-whirl-like behavior
dominates the GW- emission energetics, and $\gamma$ is a parameter
that in the geodesic analogue is related to the
instability exponent of the corresponding unstable circular
orbit.  Given the limited number of points we have, we choose to use $\gamma$ 
as our single fitting parameter.  We set $r_c$ to the average of the closest
sub and supercritical values of $r_p$ from the simulations.  We choose
$\Delta=3$ by inspection so that values of $r_p$ outside
this range are well-approximated by the NQA model. Certainly,
in future work once more data points are available it would
be preferable to more systematically fit to the other parameters
as well, and refine the model.

The top panel of Fig.~\ref{egw_fit} shows this fitting performed
individually to each set $(e,a)$ of simulations run. 
We should note that due to limited computational resources
we have not been able to perform this fitting at multiple
resolutions in order to be able to estimate the effect of truncation error on 
$\gamma$.  However, due to the sensitivity of $E_{\rm GW}$ on $r_p$
that can be seen in Fig.~\ref{egw_fit}, one would expect that the
truncation error in these values will be dominated by the resolution 
dependence of $r_c$ as opposed to the error in $\gamma$.  
In the remainder
of this section we will use the data from the $e\approx1$, $a=0$ set
to calibrate geodesic-inspired extensions to arbitrary values
of $(e,a)$, which can then be compared to the other three sets
of simulation data (bottom panel of Fig.~\ref{egw_fit}) to test how
well this extrapolation works.

To extend this model to different values of eccentricity $e$ and BH spin $a$,
we assume that the dependency of $r_c$ and $\gamma$ on these values is the same
as in the geodesic case on a Kerr background.  We will use $a_{\rm eff}$ 
to refer to the effective spin parameter we use in the Kerr formulas.
For better correspondence with the Kerr spacetime, we want to take $a_{\rm eff}$
to be the approximate spin of the BH that would form if a merger occurred.
We do this rather then use the initial spin of the BH based on results from~\cite{Pretorius:2007jn},
which suggest that far from the geodesic limit the total angular momentum 
of the binary is more important than the BH spin.  
We estimate $a_{\rm eff}$ using
\begin{equation}\label{a_eff}
 a_{\rm eff} = a_{0}\frac{\sqrt{(1+e)r_p}}{\sqrt{2r_{c0}}} + a (M_{\rm BH}/M)^2 .
\end{equation}
where $M_{\rm BH}$, $a$, $r_p$, and $e$ are the initial BH-mass, initial BH-spin, periapsis, and eccentricity, respectively, of the encounter
for which we want to compute an $a_{\rm eff}$, and
$a_{0}\approx0.5$ is the final spin measured from the $e\approx1$, $a=0$,
$r_p=r_{c 0}\approx 6.9$ simulation.  In Table~\ref{merger_bh}, we show how $a_{\rm eff}$
compares to the final BH-spins in the simulations that we followed through merger.
This simple formula based on the total angular momentum of the system
does not attempt to capture any of the complications due to differences in matter dynamics or
gravitational radiation between the different cases, and as can be seen does not always capture
the trends with $r_p$ seen in the simulations.  We use $a_{\rm eff}$ because it has a simple motivation,
and it does a decent job of estimating the final BH-spin for the purposes of this model. This 
allows us to extend this model to other values eccentricity $e$ and BH-spin $a$ that were not
simulated.

Recall that for equatorial geodesics with eccentricity $e$ 
in Boyer-Lindquist coordinates with BH-spin parameter $a$, there is a value for
the periapsis that corresponds to a marginally unstable orbit $r_c^{\rm BL}(e,a)$.  It can be found by solving the equation (see, for example,~\cite{Glampedakis}) 
$(r_c^{\rm BL})^2 = (J-aE)^2(1+e)/(3-e)$, 
where $E$ and $J$ are the orbit's specific energy and angular momentum, respectively.
For our model, we assume that
$r_c(e,a_{\rm eff}) \propto r_c^{\rm BL}(e,a_{\rm eff}) $, and we fit the proportionality constant using our numerical
results for $e\approx1$, $a=0$.

The instability exponent for unstable circular orbits in
Boyer-Lindquist coordinates is given by~\cite{Cornish:2003ig,Pretorius:2007jn}
\begin{eqnarray}
& & \gamma^{\rm BL}(e,a) = \\
& & \frac{r}{2 \pi}\left[3r^2 D+\frac{4M}{\omega^2}(rR^2\omega^2-4Ma\omega-r+2m)\right]^{-1/2}\nonumber
\end{eqnarray}
where $R=r^2+a^2(1+2M/r)$, $D=r^2+a^2-2Mr$, $\omega=M/(Ma\pm \sqrt{Mr^3})$ ($\pm$ for prograde and retrograde, respectively),
and we set $r=r_c^{\rm BL}(e,a)$. 
Again, we assume that
$\gamma(e,a_{\rm eff}) \propto \gamma^{\rm BL}(e,a_{\rm eff}) $ and fit the proportionality constant using our numerical
results for $e\approx1$, $a=0$.
For the angular momentum lost to gravitational waves in a close-encounter we assume a similar expression
\begin{equation}
J_{\rm GW}=J_0(1-(\delta r_p/\Delta)^\gamma)
\end{equation}
where again $J_0$ is the difference in angular momentum of a quasicircular orbit and an 
eccentricity $e$ orbit evaluated at the same separation $r_p$.

This simple prescription for estimating gravitational-wave energy and angular-momentum 
loss has some obvious limitations.  We are extrapolating the critical impact parameter
and instability exponent based on a Kerr spacetime to a BH-NS spacetime that is dynamic
and nonperturbative.  We are also ignoring tidal effects or dependence on EOS in this model. 
Nevertheless, in Fig.~\ref{egw_fit}, we compare how closely this predicted scaling with spin and 
eccentricity matches that from simulation results.  Given its simplicity,
as well as the numerical error in these results, this model does a satisfactory job of capturing
the trends in these scalings.

\subsection{Systems undergoing multiple close-encounters}
\begin{figure*}
\begin{center}
\includegraphics[width=2.30in,clip=true,draft=false]{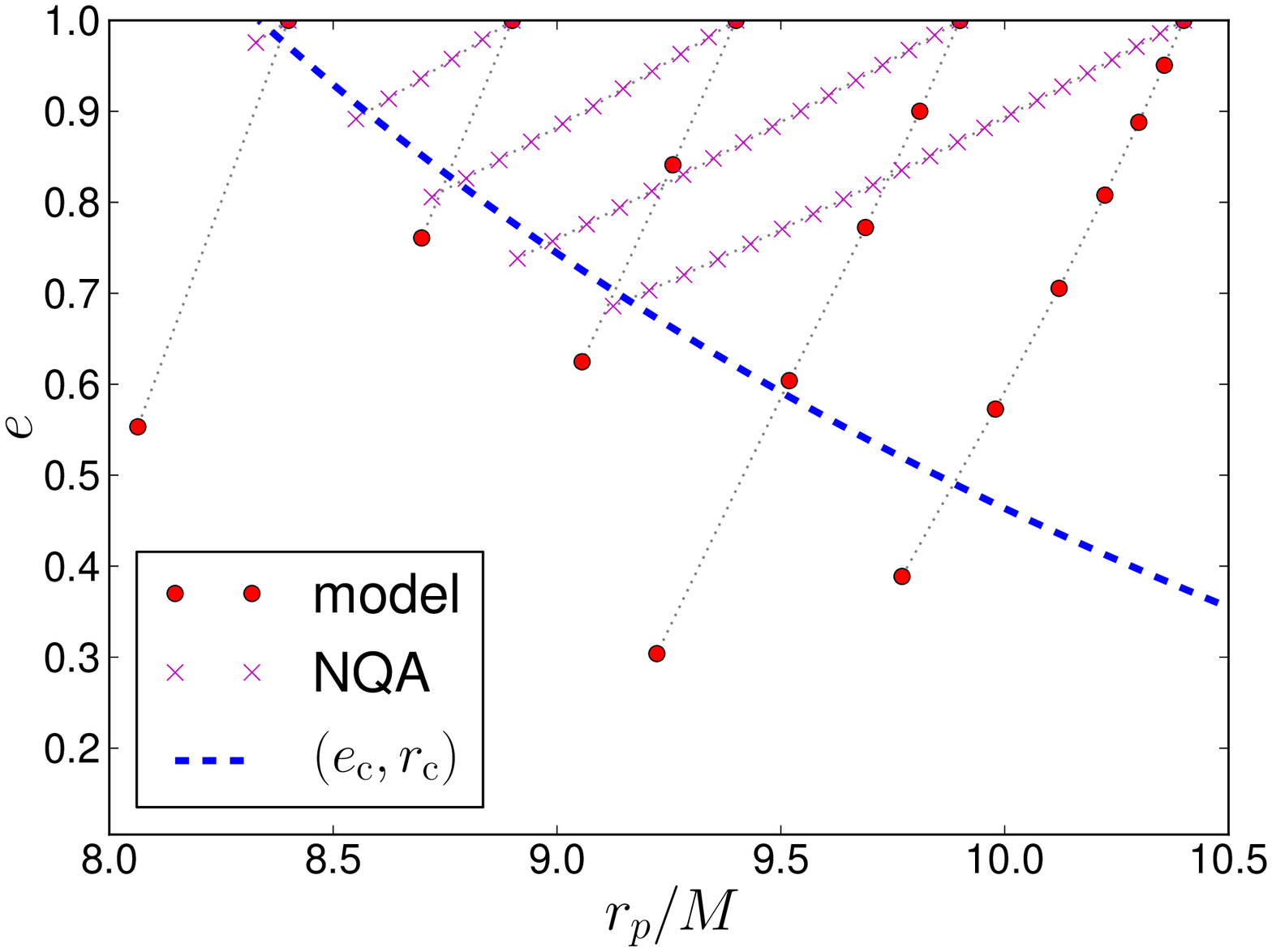}
\includegraphics[width=2.30in,clip=true,draft=false]{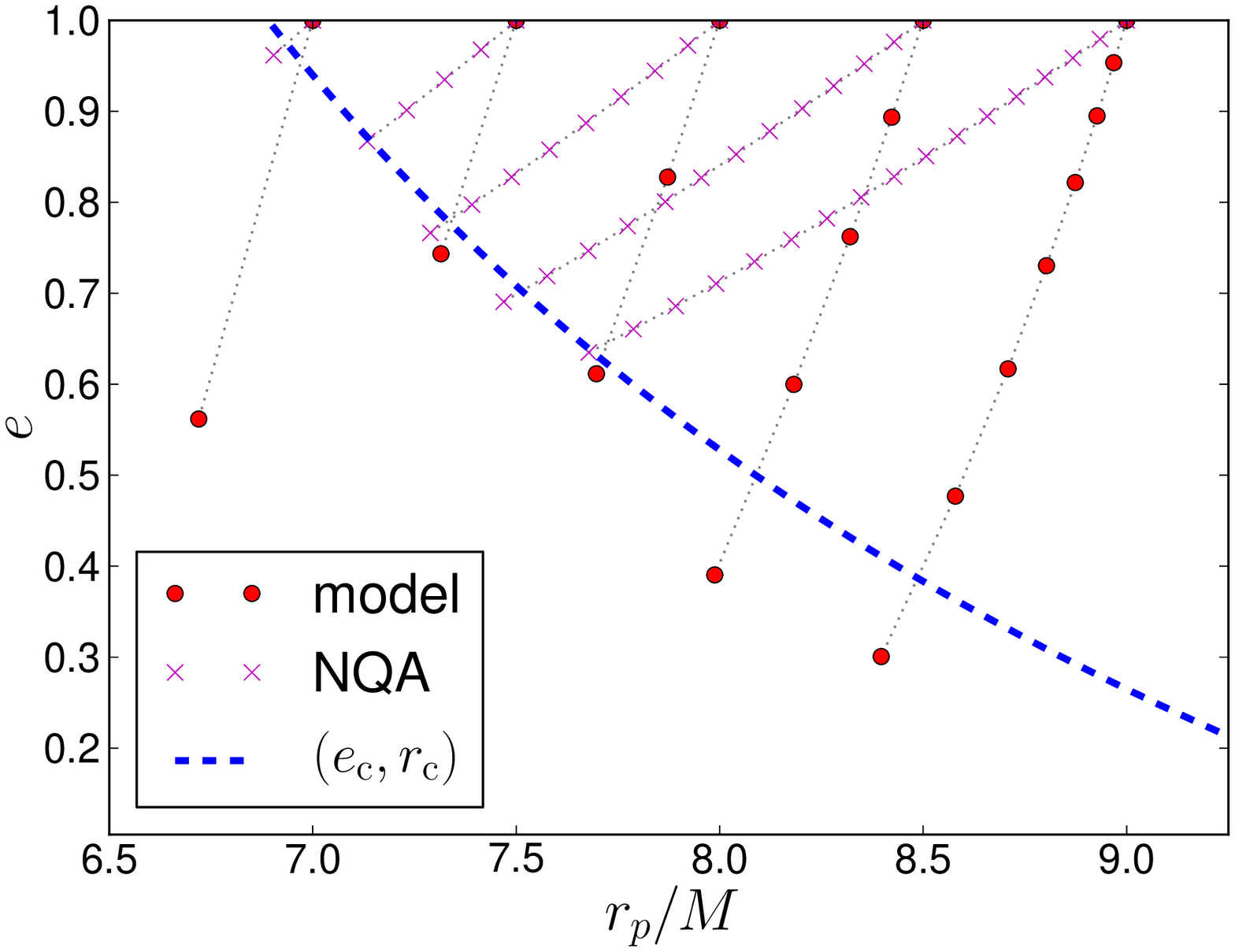}
\includegraphics[width=2.30in,clip=true,draft=false]{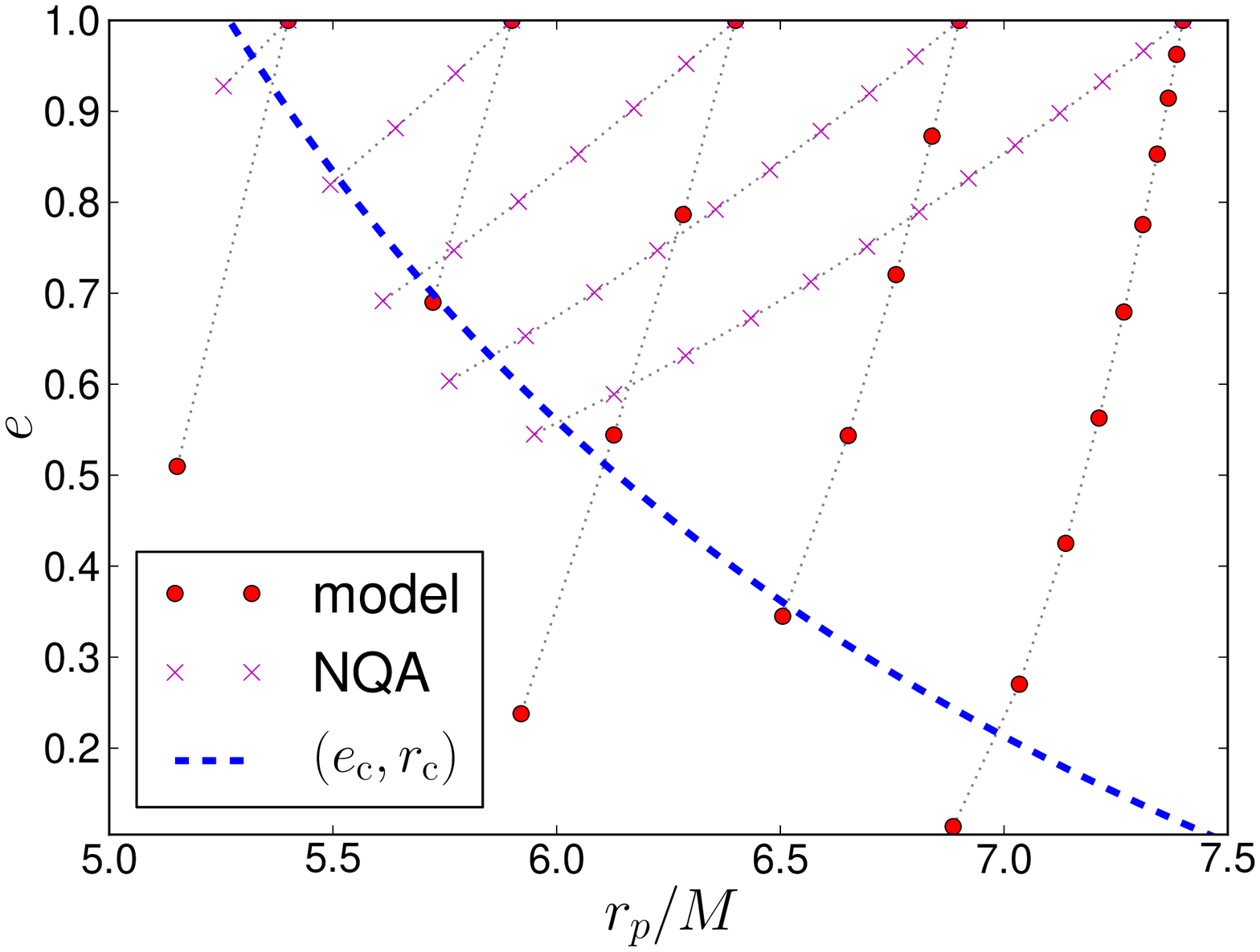}
\caption
{
The evolution of the eccentricity and periapsis separation of various $4:1$ 
mass ratio BH-NS binaries that begin marginally unbound and undergo a 
series of close-encounters (large red points) before merging.  For 
comparison, we also plot the results using the NQA expressions 
from~\cite{pm63,turner77} (magenta x's).  We also plot the 
critical eccentricity for a given $r_p$ for a close-encounter to result in merger (blue, dotted-line).
Hence, the points below this curve correspond to merger events.
From left to right the plots 
correspond to an initial BH-spin of $a=-0.5$, 0, and 0.5, respectively.
}
\label{orbit_evo}
\end{center}
\end{figure*}

The enhanced GW energy and angular-momentum losses during a close-encounter 
for a given $r_p$ result in more rapid loss of 
eccentricity and larger rate of decrease of $r_p$ of the next encounter. Figure~\ref{orbit_evo} shows approximate trajectories in $e$ 
and $r_p$ for binaries on initially marginally unbound ({\em i.e.} $e=1+\epsilon$)
orbits for a range of initial BH spin.  These results 
were obtained using the above model for energy and angular momentum lost to GWs
and assuming $e$ and $r_p$ follow the Newtonian relationship to energy and angular momentum.
We approximate these losses as occurring in discrete steps during close-encounters. (This approximation
will break down as $e\rightarrow 0$.) 
Trajectories computed with the 
NQA amplitude and eccentricity-dependence~\cite{pm63,turner77,Berry:2010gt} are also shown in the figure.  
The NQA execute many more orbits before merger.  This clear departure from
the NQA prediction at small $r_p$ due to strong field-GR effects should 
thus be apparent in the gravitational waveform.  
This model can also be used to predict approximately
the critical-initial impact parameter for the BH-NS system to merge on the second close-encounter, the 
third close-encounter, and so on.  We expect interesting dynamics around each of the thresholds up until the point
where the system has undergone enough close-encounters that it circularizes.

\section{Conclusion}\label{conclusions}
We have performed an initial survey of eccentric BH-NS mergers including
the effects of black hole spin and varying the NS EOS.  Though the limited
number of values we considered in this
work does not begin to exhaust the parameter space, what is immediately apparent
is the strong diversity in the resulting gravitational and matter dynamics.  
Though we have not yet studied the consequences of this on gravitational-wave 
detectability and parameter extraction, or possible electromagnetic 
counterparts, it is clear that the outcome can depend sensitively on the 
binary parameters and matter EOS. 

In order to understand the effects of 
large eccentricity, these simulations can be compared to
the quasicircular BH-NS merger simulations of~\cite{shibataBHNS3,Kyutoku:2011vz}, 
which used the same piecewise polytrope equations of state.  In~\cite{shibataBHNS3}, 
it was found that for a nonspinning BH and a 3:1-mass ratio, the resulting disk masses 
where 0.044, 0.0015, and $<10^{-5}$ $M_{\odot}$ for the 2H, HB, and B EOSs, respectively.
Hence eccentric mergers with certain ranges of impact 
parameters is one way to achieve significantly larger accretion disks compared to the 
quasicircular case without BH spin.  The effects of BH spin were considered 
in~\cite{Kyutoku:2011vz}, where for a 4:1-mass ratio and the HB EOS they found disk 
masses of 0.024 and 0.18 $M_{\odot}$ for BH spins of $a=0.5$ and 0.75, respectively 
(and even larger values for stiffer EOSs or lower mass ratios).  They also only
found a non-negligible amount of unbound material ($\geq 0.01 M_{\odot}$) only for stiff EOSs
like the 2H.  Thus, an important characteristic of eccentric mergers is the larger
amount of ejected material found for some parameters. This could be a significant
source of r-process 
elements~\cite{1974ApJ...192L.145L,Rosswog:1998gc,Li:1998bw}, and 
give rise to EM counterparts, {\em e.g.} 
through nuclear decay of the radioactive r-process isotopes~\cite{2010MNRAS.406.2650M,Metzger:2011bv}.
The amount of energy radiated away in gravitational waves for the quasicircular case is somewhat
higher than the values found here (for example, $E_{GW}= 1.7\%$ of the total mass for a 
4:1 mass ratio, HB EOS, and $a=0.5$ in~\cite{Kyutoku:2011vz}).  The gravitational-wave signal is 
of course completely different for eccentric mergers, which are dominated by bursts 
from close-encounters or merger of the two compact objects.    
 
Whether dynamical capture BH-NS mergers occur
with sufficient frequency in the Universe to constitute a decent event rate for
ground-based GW detectors is another question; at the least, failure to observe
such events will place constraints on these sources, while if several are observed
given the sensitivity of the outcome to properties of the binary, they could be ideal
environments to reveal the structure of neutron stars.
And again, we should emphasize that excessive fine tuning of parameters is
not required for significant variability. Taking the relative velocity
at large separations in a typical nuclear-cluster cusp as an example ($\approx 1000$~km/s),
roughly $25\%$ of such encounters will have $r_p\lesssim 10$, (following the same
line of reasoning discussed in the introduction for the percentage of direct
collisions) corresponding to the cases
studied here. Certainly some of the most extreme examples of mass
transfer, large accretion disks, or multiple whirl orbits will be rare;
perusing Tables ~\ref{master_table}-\ref{bound_table} will give
some idea of the distribution with $r_p$. (Recall that the cross section scales linearly 
with $r_p$ due to gravitational focusing, and that one would also expect some cases
with larger initial $r_p$ than we followed through merger to exhibit
similar variability.)
It is not trivial to calculate a transition value of initial $r_p$ 
above which the qualitative behavior at late times is described by a quasicircular
inspiral, though our study suggests at least a quarter of dynamical-capture 
binaries will merge with high eccentricity.

For future work, we plan to investigate the detectability of the GW signals
from BH-NS mergers that arise from dynamical capture in the strong-field 
regime. This will complement the first study of such systems presented 
in~\cite{Kocsis:2011jy}, in that we intend to focus on the later stages of
high-eccentricity mergers, including the merger/ringdown part of the signal.
Future work also includes expanding the parameter space to different 
BH and NS masses, BH-spin orientations, and (as computational resources permit)
evolution of systems that exhibit more than two close-encounters before
merger. Performing higher-resolutions simulations will also be important
to coming up with a more quantitatively accurate model of the behavior 
of the BH-NS binary near the threshold for merger during a given close-encounter.
Doing so will not only require additional computational resources but 
a better method for creating initial data describing such eccentric binaries,
since the superposition method used here imposes an effective floor on 
the accuracy that be reached.
It would also certainly be interesting to investigate EM counterparts to these
events; such simulations would require extensions to the code used here
beyond the present GRHD.

\acknowledgments
We thank Adam Burrows, John Friedmann, Roman Gold, Janna Levin, Charalampos Markakis, 
Benjamin Lackey, Sean McWilliams and 
Richard O'Shaughnessy for useful conversations.  
This research was supported by the NSF
through TeraGrid resources provided by NICS under Grant No.
TG-PHY100053, the Bradley Program fellowship (BCS), the NSF 
Graduate Research Fellowship under Grant No. DGE-0646086 (WE), NSF Grant No. 
PHY-0745779 (FP) and Grant No. PHY-1001515 (BCS),
and the Alfred P. Sloan Foundation (FP). 
Simulations were also run on the {\bf Woodhen} cluster 
at Princeton University.

\bibliographystyle{unsrt}
\bibliography{bhns}

\end{document}